\documentclass{article}

\usepackage[english]{babel}
\usepackage[letterpaper,top=2cm,bottom=2cm,left=3cm,right=3cm,marginparwidth=1.75cm]{geometry}
\usepackage{mathptmx}
\usepackage{sectsty}
\usepackage{amsmath}
\usepackage{bm}
\usepackage{graphicx} 
\usepackage{hyperref}

\usepackage{subcaption}

\usepackage[format=plain,justification=justified,singlelinecheck=false,font={stretch=1.125,small,sf},labelfont=bf,labelsep=space]{caption}

\usepackage{newtxmath}

\title{Sedimenting rigid particles of certain shapes approach a stationary orientation}

\author{
    Chandra Shekhar\textit{$^{a}$}$^\|$,  
    Harish N. Mirajkar\textit{$^{a}$}$^{\S}$,
    Piotr Zdybel\textit{$^{a}$},
    Yevgen Melikhov\textit{$^{a}$},
    \and Maria L. Ekiel-Je\.zewska\textit{$^{a}$}$^{\ast}$
}

\begin{document}

\renewcommand{\thefootnote}{\fnsymbol{footnote}}
\renewcommand\footnoterule{\vspace*{1pt}
 \hrule width 3.5in height 0.4pt \vspace*{5pt}} 
\newcommand{\bee}{\begin{eqnarray}}
\newcommand{\eee}{\end{eqnarray}}
\newcommand{\ba}{\begin{array}}
\newcommand{\ea}{\end{array}}
\newcommand{\bt}{\begin{tabular}}
\newcommand{\et}{\end{tabular}}

\maketitle

\begin{abstract}
This work investigates experimentally and numerically the dynamics of rigid particles settling under gravity in a highly viscous fluid. We demonstrate that certain shapes: cones, crescent moons, arrowheads, and open flat rings reorient and approach a stationary configuration. We determine the mobility coefficients and the characteristic reorientation times. We find out that the two rotational-translational mobility coefficients have opposite signs. Therefore, based on the equations of motion for rigid bodies with two orthogonal planes of symmetry, theoretically derived in Joshi \textit{and} Govindarajan (\textit{Phys. Rev. Lett.}, vol.~134, 2025, 014002) and Ekiel-Je\.zewska \textit{and} Wajnryb (\textit{J. Phys. Condens. Matter}, vol.~21, 2009, 204102), we conclude that the approached stationary configurations are stable. Owing to the similarity principle, our experimental findings apply to microobjects in water-based solutions. The reorientation of sedimenting rigid particles of certain shapes to a stationary stable configuration in a relatively short time might be used for biological, medical, or industrial applications.
\end{abstract}

\footnotetext{\textit{$^{a}$}~Institute of Fundamental Technological Research, Polish Academy of Sciences, ul. Pawi\'nskiego 5B, 02-106 Warsaw, Poland}

\footnotetext{$^\|$~Chandra Shekhar, Piotr Zdybel and Yevgen Melikhov contributed equally to this work}

\footnotetext{$^{\S}$~Currently at School of Engineering and Applied Science, Ahmedabad University, Gujarat, India}

\footnotetext{$^{\ast}$~corresponding author: Maria L. Ekiel-Je\.zewska E-mail: mekiel@ippt.pan.pl}

\footnotetext{\dag~Author ORCIDs: 
    Chandra Shekhar, https://orcid.org/0000-0002-4665-2915; 
    Harish N. Mirajkar, https://orcid.org/0000-0002-7365-9514; 
    Piotr Zdybel, https://orcid.org/0000-0001-7484-1425; 
    Yevgen Melikhov, https://orcid.org/0000-0002-9787-5238; 
    Maria L. Ekiel-Je\.zewska, https://orcid.org/0000-0003-3134-460X.
}

\section{Introduction}
Sedimentation of single rigid particles of different complex shapes in a viscous fluid for the Reynolds number much smaller than unity has been studied for a very long time (see early papers \cite{brenner1963stokes,brenner1964stokes}). One of the basic tasks, studied experimentally and numerically, e.g., in \cite{lasso1986stokes,cichocki1995stokes,weber2013sedimentation} for hollow cylinders, conglomerates of spheres, and rigid knots, is to determine how much the particle's Stokes settling velocity is smaller than that of a sphere of equal mass and volume. In general, the settling velocity of elongated particles or disks depends on their orientation and is accompanied by a lateral drift \cite[p. 605]{lamb1924hydrodynamics}; \cite{taylor1966low}. Moreover, depending on shape, a particle can rotate while settling.

For the incompressible fluid flow satisfying the stationary Stokes equations, the particle angular velocity and the translational velocity of its centre-of-mass are determined by the $3\times 3$ rotational-translational and translational-translational mobility matrices, respectively, applied to the gravitational force, as explained, e.g., in \cite{happel1965low,durlofsky1987dynamic,felderhof1988many,kim,Graham}. It is known that for chiral particles, there exists a rotational-translational coupling, which means that the rotational-translational mobility matrix does not vanish. Coupled rotational and translational dynamics of chiral objects have been extensively studied, including rigid knots \cite{gonzalez2004dynamics}, chiral propeller-like shapes \cite{doi2005motion},  asymmetric rigid fibres~\cite{tozzi2011settling,vahey20113d}, rigid  helices~\cite{palusa2018sedimentation}, helical rigid ribbons~\cite{huseby2025helical}. However, the rotational-translational coupling, investigated, e.g., in \cite{gonzalez2004dynamics,doi2005sedimentation}, can be non-zero also for particles with a non-chiral shape. Examples are rigid non-chiral propellers considered in \cite[p. 119]{kim}; \cite{makino2003sedimentation,doi2005sedimentation,krapf2009chiral},
rigid trumbbells investigated numerically and theoretically in~\cite{ekiel2009hydrodynamic}, and bent rigid disks, studied experimentally, numerically, and theoretically in~\cite{miara2024dynamics,vaquero2024u,vaquero2025fluttering}. The non-chiral propellers, trumbbells, and bent disks have two perpendicular symmetry planes.

Sedimentation dynamics of bodies with two orthogonal planes of symmetry have been recently determined~\cite{joshi}. 
Namely, it was shown there that each shape belongs to one of three distinct families of solutions: settlers, which tend to a stationary stable orientation characterised by vertical terminal velocity; drifters, which approach an inclined orientation with a lateral drift superimposed with a vertical translation; and flutterers, which perform periodic rotations while sedimenting.

In this paper, our task is to provide experimental and numerical examples of the settler dynamics, and demonstrate that the typical times to approach the stationary stable orientation are short enough for potential applications.
We are going to study experimentally, numerically, and theoretically the sedimentation of single rigid micro-objects (in the following called particles) of different shapes and determine how they translate and rotate under gravity in a viscous fluid. Typically, the characteristic Reynolds number for the motion is much smaller than unity, $\mathrm{Re} \ll 1$. The fluid and particle motion in a geometrically and hydrodynamically similar system, but with larger objects and more viscous fluid, will be scaled just by the similarity factor. "Hydrodynamically similar" means that the Reynolds number for both systems is the same. Here, the Reynolds number 
\bee&&
    \text{Re}= \frac{\rho U_f L}{\eta}
\eee
is equal to the fluid density $\rho$ times a particle velocity $U_f$ times a particle characteristic dimension $L$ divided by the fluid dynamic viscosity $\eta$. 
In our experiments, we employ the similarity principle, studying the dynamics of objects approximately the size of a few millimetres that settle in a very viscous silicon oil. Typically, the Reynolds number in the experiments is within the range  $\mathrm{Re}~= 2.1 \cdot 10^{-3} - 1.1\cdot 10^{-2}$.

The structure of the paper is the following. In section~\ref{thb}, we provide theoretical background and explain the characteristic feature of the rotational-translational mobility matrix for the settlers. In section~\ref{exp}, the experimental setup, the particles, and the methods are described. Section~\ref{eresults} presents the experimental results: typical examples of the translation and rotation of particles, shown by the time sequence of snapshots, the time-dependent vertical velocity, and the time-dependent inclination angle, as well as the average values of the mobility coefficients. In section~\ref{tresults}, the experimental shapes are approximated as conglomerates of identical spherical beads, and their mobility coefficients are determined numerically. Section~\ref{conclusions} contains the comparison of the theoretical and experimental results, together with a theoretical explanation of the instabilities seen in the experiments, and the conclusions. In Appendix~\ref{sec:other_behaviour}, we present more examples of the particle reorientation. Appendix~\ref{number} contains statistics of all the experimental trials. Appendix~\ref{extracting} explains the method used to determine the time-dependent inclination angle $\theta$. Appendix~\ref{sec:experimental_parameters} contains the fit parameters for $\tan(\theta/2)$ as functions of time for exemplary experimental trials. Finally, Appendix~\ref{sec:theor_param} describes the construction of the bead model for the particles of different shapes used in our experiments.

\section{Theoretical background}\label{thb}

We consider a rigid particle moving in a viscous fluid under the influence of an external force  $\bm{F}$, equal to gravity minus buoyancy. 
We assume that the Reynolds number is much smaller than unity,
 Re $\ll$ 1, and that the fluid velocity $\bm{v}(\bm{r})$ and pressure $p(\bm{r})$ are described by the stationary Stokes equations, 
\begin{equation}
    \eta \bm{\nabla}^2 \bm{v} - \bm{\nabla} p=\bm{0},~~~~~ \bm{\nabla}\cdot \bm{v}=0.
    \label{stokes}
\end{equation}
The motion of the particle is derived from equation~\eqref{stokes}, assuming the stick boundary conditions for the fluid at the particle surface \cite{kim,Graham}. The particle translational and angular velocities, $\bm{U}$ and $\bm{\Omega}$, are linear functions of the Cartesian components of the external force $\bm{F}$ 
acting on the particle. We calculate $\bm{U}$ with respect to the centre of mass. The benefit is that the torque vanishes with respect to the centre of mass, and therefore,
\bee
    &&\left(\ba{l}\bm{U}\\
    \bm{\Omega}\ea\right) = \left(\ba{ll}\bm{\mu}^{tt} \;\bm{\mu}^{tr}\\
    \bm{\mu}^{rt} \; \bm{\mu}^{rr}\ea\right) \cdot \left(\ba{l}\bm{F}\\
    \bm{0}\ea\right),
\eee
or alternatively,
\bee
    &&\bm{U}= \bm{\mu}^{tt} 
    \cdot \bm{F},\hspace{1cm}
    \bm{\Omega} = \bm{\mu}^{rt} \cdot \bm{F}.
\eee
The $3 \times 3$ Cartesian matrices: the translational-translational mobility $\bm{\mu}^{tt}$ and the rotational-translational mobility $\bm{\mu}^{rt}$ depend on the particle orientation.

In this paper, as \cite{joshi} and \cite{ekiel2009hydrodynamic} in their theoretical studies, we focus on the particles having two perpendicular symmetry planes (but not symmetric with respect to reflection in the third perpendicular plane). We choose the frame of reference with the coordinates $x$ and $y$ perpendicular to the symmetry planes, 
and with the $x$-coordinate along $L$ - the larger from $x$ and $y$ sizes of the particle.
(Most of our particles are rather flat - their size along $y$ is smaller than along $x$; for the cone, exceptionally, $x$ and $y$ sizes are the same.)
In this special frame of reference, the dimensionless mobility matrices have the form,
\bee
    &&\pi \eta L \bm{\mu}^{tt}=\left(\bt{ccc}$\mu_{11}$ &0 &0\\
    0 &$\mu_{22}$ &0\\
    0& 0 &$\mu_{33}$\et \right),
    \hspace{1cm}
    \pi \eta L^2 \bm{\mu}^{rt} = \left(\bt{lll} 0 & $\mu_{42}$ &  0\\
    $\mu_{51}$ & 0 & 0\\
    0& 0 &0\et
    \right).
    \label{matrices}
\eee

In the following, we will use $L$ as the length unit, and the dimensionless time $t$ defined  as
\bee
    t=\frac{F}{\pi \eta L^2}\tilde{t},
    \label{eq:time}
\eee
with $F=|\bm{F}|$ and $\tilde{t}$ denoting the dimensional time.

\begin{figure}
    \centering
    \begin{subfigure}{0.34\textwidth}
        \centering
        \includegraphics[width=1.0\textwidth]{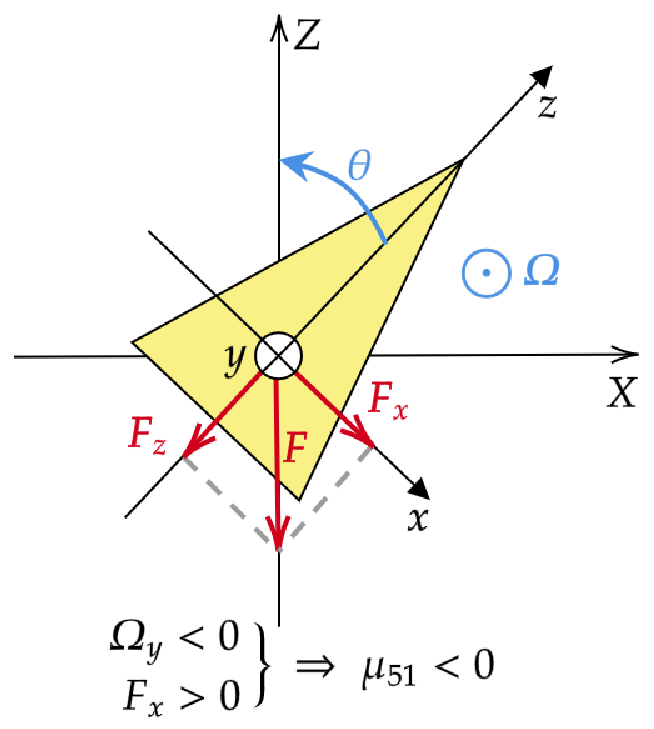}
        \caption{}
        \label{fig:sing_conv:mu51}
    \end{subfigure}
    \hfill
    \begin{subfigure}{0.34\textwidth}
        \centering
\includegraphics[width=1.0\textwidth]{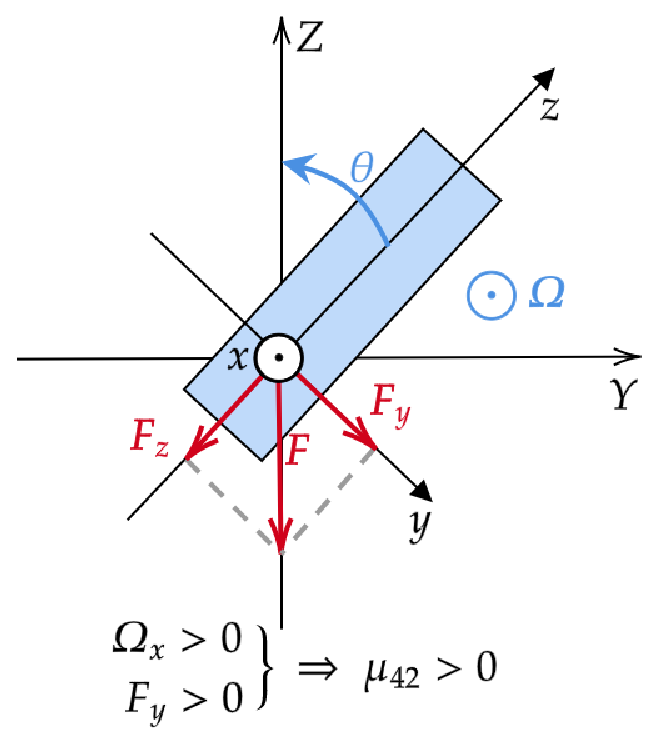}
        \caption{}
        \label{fig:sing_conv:mu42}
    \end{subfigure}
        \begin{subfigure}{0.30\textwidth}
        \centering
\includegraphics[width=0.665\textwidth]{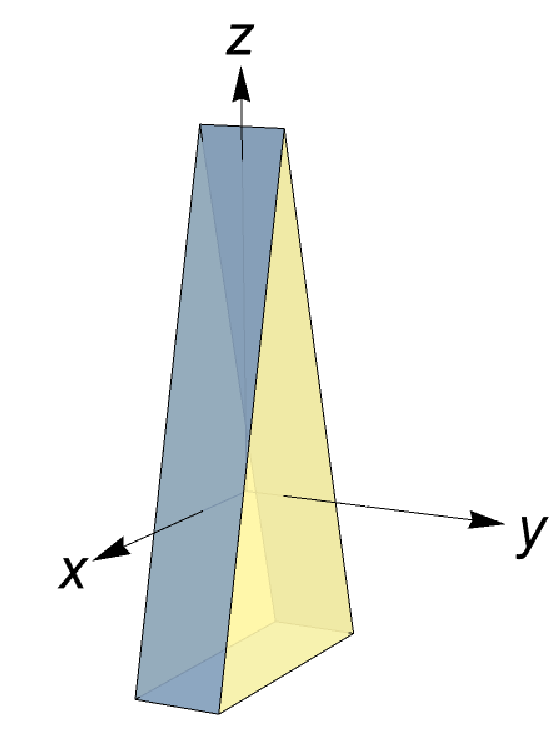}
        \vspace{1.5cm}
       \caption{}
        \label{fig:sing_conv:body}
    \end{subfigure}
    \vspace{-0.2cm}
    \caption{
    Schematic illustration depicting the notation and the interpretation of the signs of the rotational–translational elements of the mobility matrix: (a) $\mu_{51}$ and (b) $\mu_{42}$. 
    Here, $XYZ$ is the laboratory frame of reference, while $xyz$ is the body-fixed coordinate system. 
    The force $\boldsymbol{F}$ acting on the body is antiparallel to the $Z$-axis. 
    Figure (c) illustrates a 3D view of the exemplary wedge-shaped body with two orthogonal symmetry planes.}
    \label{fig:sing_conv}
\end{figure}

We consider such particles that rotate and approach a stationary configuration, 
regardless of their initial orientation. This family of shapes has been called ``settlers'' in the theoretical study \cite{joshi}. For the settlers, the rotational-translational coefficients $\mu_{42}$ and $\mu_{51}$ are different from zero and have the opposite signs,
\bee
    && \mu_{42} \;\mu_{51} < 0.
\eee
Figure~\ref{fig:sing_conv} shows the interpretation of the signs of the off-diagonal mobility coefficients: (a) $\mu_{51}$ and (b) $\mu_{42}$.

In the laboratory frame of reference $XYZ$, the total force $\bm{F}$ is antiparallel to the $Z$-axis, $\bm{F}=-F\bm{\hat{Z}}$.
If we choose the initial conditions in such a way that during the motion the axes $y$ and $Y$ are parallel to each other (as in figure~\ref{fig:sing_conv:mu51}), then, owing to symmetry,  the motion takes place in the plane perpendicular to $Y$, and the particle orientation is specified by the angle $\theta(t)$ between the axis $Z$ and $z$ which evolves with time $t$ as
\bee
 &&  \frac{\mathrm{d}\theta}{\mathrm{d}t}= \mu_{51} \sin \theta.
    \label{eq:theta_51_exp}
\eee
If we choose the initial conditions in such a way that during the motion the axes $x$ and $X$ are parallel to each other (as in figure~\ref{fig:sing_conv:mu42}), then, owing to symmetry,  the motion takes place in the plane perpendicular to $X$, and the particle orientation is specified by the angle $\theta(t)$ between the axis $Z$ and $z$ which evolves with time $t$ as
\bee
    && \frac{\mathrm{d}\theta}{\mathrm{d}t}= -\mu_{42} \sin \theta.
    \label{eq:theta_42_exp}
\eee
Here, equations~\eqref{eq:theta_51_exp} and \eqref{eq:theta_42_exp} are identical to equation (14) introduced in \cite{ekiel2009hydrodynamic} and solved there as their equation (15). Therefore, the solutions of equations~\eqref{eq:theta_51_exp} and \eqref{eq:theta_42_exp} read,
\bee
    && \tan(\theta/2)= A \exp(\mu_{51} t),\label{so51}\\
    && \tan(\theta/2)= A \exp(-\mu_{42} t).\label{so42}
\eee

The general expressions for the settler's non-planar motion are given in \cite{ekiel2009hydrodynamic} in their equations (10)-(12) and also in \cite{joshi}. To simplify the analysis, in the experiments we aimed towards providing the initial orientation of the particle in the plane perpendicular to $y$ or $x$, measuring the angle $\theta$ as a function of time, and then comparing with the corresponding solution from equations~\eqref{so51} or \eqref{so42}, as it will be described in section~\ref{eresults}. In the experiments, we will analyse directly the dimensional quantities, and in particular, the time coefficients defined as 
\bee
      \tilde{\tau}_{51} \equiv \frac{\pi \eta L^2}{F} \frac{1}{\mu_{51}},\;\; \mbox{ and }\;\;\;\; \tilde{\tau}_{42} \equiv \frac{\pi \eta L^2}{F} \frac{1}{\mu_{42}}.
    \label{eq:mobility}
\eee
from the requirements that $\mu_{51}t=\tilde{t}/\tilde{\tau}_{51}$ and $\mu_{42}t=\tilde{t}/\tilde{\tau}_{42}$.

In the experiments, we will show that particles reorient and reach the stationary configuration (see section~\ref{eresults}). By measuring the final vertical stationary centre-of-mass velocity $U_{\!f}$, and the total force $F$, we will determine the dimensionless translational-translational mobility coefficient relevant for the particle translation in the stationary state, 
\bee
    && \mu_{33} = \pi \eta L \frac{U_{\!f}}{F}.
    \label{eq:mju33}
\eee

Each of the particle shapes used in our experiments is later approximated as a collection of touching beads, and the mobility coefficients in equation~\eqref{matrices} are evaluated by the multipole method corrected for lubrication, outlined in \cite{cichocki, ekiel-jezewska2009} (see section~\ref{tresults} for the results).

\section{Experimental system, materials, and methods}\label{exp}
\subsection{Setup}
\begin{figure}[h!]
    \centering
    \includegraphics[width=0.84\linewidth]{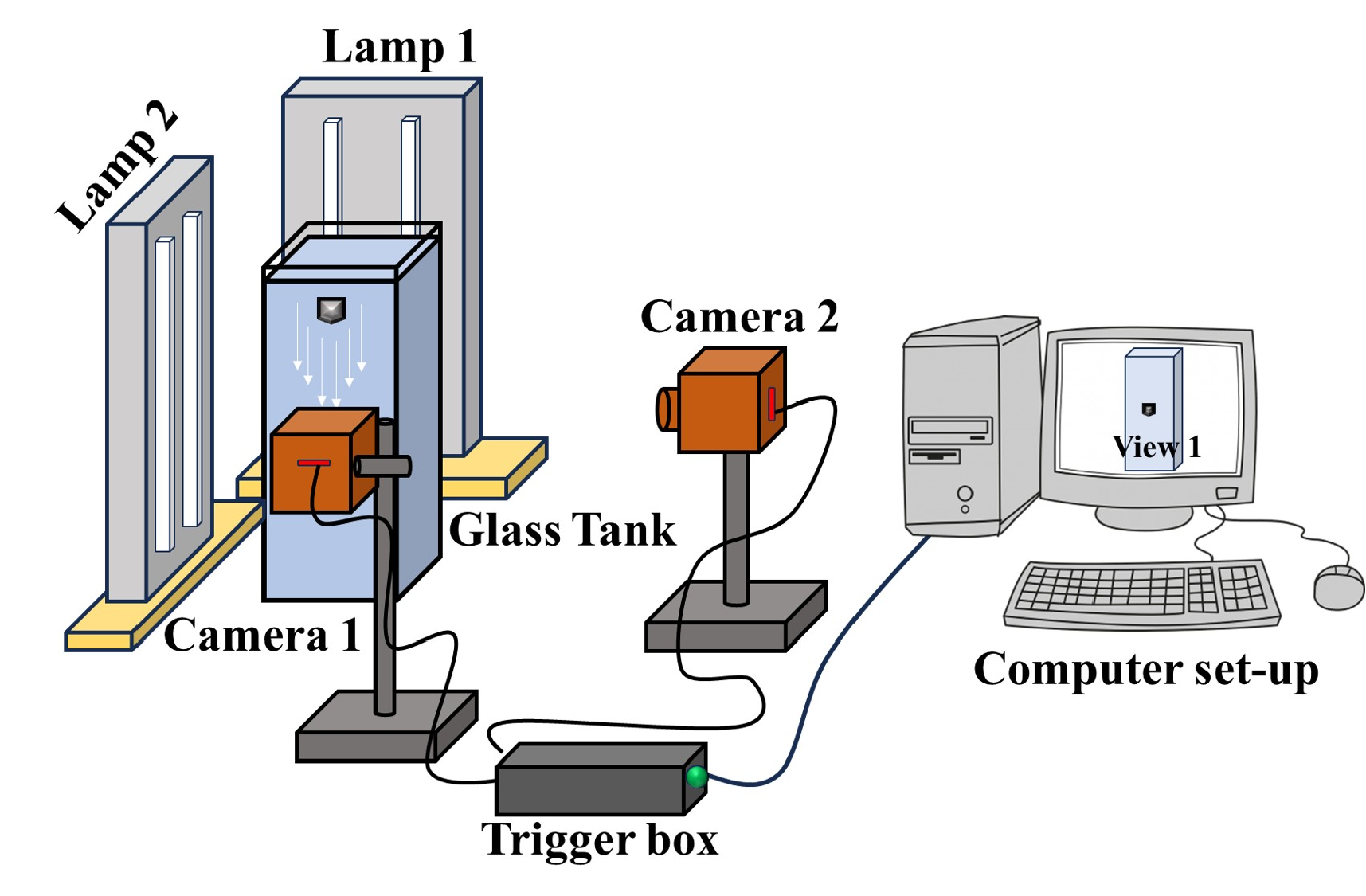}
    \vspace{-0.3cm}
    \caption{
    Schematic representation of the experimental setup.
    }
    \label{setup}
\end{figure}

The setup is almost the same as the one used in \cite{shashank_dynamics_2023}, but without a mirror. 
The experiments are conducted in the glass tank of inner dimension 200 mm $\times$ 200 mm $\times$ 500 mm (width, depth, and height, respectively). We utilize the highly viscous silicone oil (manufactured by Silikony Polskie) with the kinematic viscosity $\nu = 5 \cdot 10^{-3} \text{ } \text{m}^2 \text{ s}^{-1}$ and density $\rho = 970 \text{ } \text{kg m}^{-3}$ at 25$^\circ$C. 
Solid objects (in the following called particles) with two orthogonal planes of symmetry are dropped in this tank to study if they reorient and approach a stationary configuration. 
Particle motion is recorded using two cameras placed perpendicularly to the front and side of the tank, with the line of sight directed to the centre of the wall. 
Both cameras look at the glass tank directly without any obstacles, and both lamps were kept parallel to the walls of the tank with the diffusers to get equal illumination throughout the tank area, as shown in figure~\ref{setup}. 
The arrangement of the light source, tank, and cameras ensures that the opaque sedimenting particles are clearly visible on a bright background.

We use two identical full-frame DSLR cameras (Canon 5D Mark IV) with a resolution of 30 megapixels,  equipped with 100 mm prime lenses. The cameras are connected to a multiple-camera operation trigger box (manufactured by Esper Ltd., U.K.). The cameras are triggered simultaneously using the Esper trigger box operating software, controlled by the computer setup. The time delay of the trigger box for triggering both cameras simultaneously is 20 ms. 
The images are captured at the rate of 1 frame per second. 
Both cameras were set with the highest available $f$-number, $f=32$ (corresponding to the smallest aperture) and with an exposure time of 1/25 second.  
 
The distance travelled by the object in the given exposure time was in the range of 1-7 pixels.
The motion of the object during the exposure time can be seen as blurring of the image. The influence of the blurring on extracting values of the angle $\theta$ is analysed in Appendix~\ref{extracting}, with the conclusion that the effect of the motion blur can be safely neglected in the present study.
The ISO of both cameras is kept at 160 to achieve sufficient brightness with a moderate noise level.  

Both cameras are arranged in portrait orientation, positioned equidistant from the front and side faces of the tank at approximately 820 mm. 
This distance has been selected to ensure the complete field of view captured by each camera, covering slightly more than 300 mm vertically and 200 mm horizontally across the height and width of the tank, respectively. 
Both cameras are aligned at the tank’s mid-height. 
This setup avoids recording the particle motion closer than $\sim$100 mm from the top or bottom of the fluid to avoid recording a particle's strong hydrodynamic interaction with the free surface and the bottom wall. 
Note that all the settings in both cameras are identical as described in the above section.

\subsection{Rigid particles}\label{sec:exp_sys:rigid_particles}

The study involves rigid objects (sourced from Koniarscy S. C., Poland) of different shapes with two perpendicular symmetry planes, non-symmetric with respect to reflections in the third perpendicular plane: 
crescent moons 
and cones 
made of glass, 
arrowheads fabricated from hematite (non-magnetic), 
and open rings made of steel. 
We opened the rings manually, adjusting the size of the gap between the ends. 

\begin{figure}[b!]
    \centering
    \includegraphics[width=15.7cm]{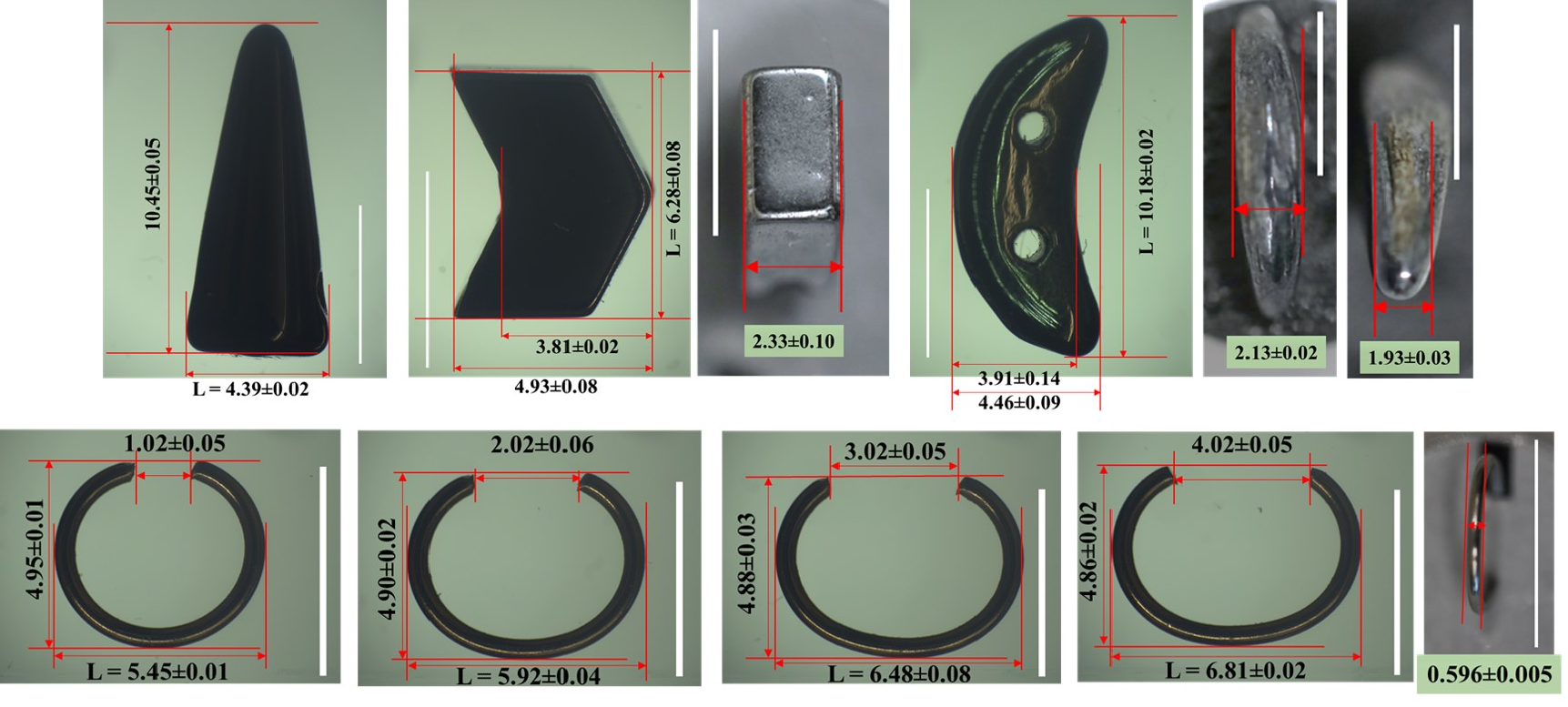}
    \vspace{-0.2cm}
    \caption{
    The objects used in the experimental studies, shown from different perspectives, with their dimensions indicated in millimetres (mm). The scale bar (white line) represents 5~mm. 
    }
    \label{objects}
\end{figure}

Figure~\ref{objects} presents the real images of the objects used in this study. 
All dimensions shown in figure~\ref{objects} were measured using a high-precision digital vernier calliper with an accuracy of 0.01 mm per measurement. Each measurement was performed on at least five objects and repeated 2–3 times to take into account that the dimensions of individual objects may vary, and to minimise measurement errors. 
The dimensions of the objects and their standard deviations are as follows:
\begin{enumerate}
    \item \textbf{Cone-shaped objects:} 
        The diameter is $L= 4.39 \pm 0.02$ mm, and 
        the height is $H=10.45 \pm 0.05$ mm.
    \item \textbf{Arrowhead-shaped objects:} 
        The length is $L=6.28 \pm 0.08$ mm, 
        the total height is $H=4.93 \pm 0.08$ mm, 
        the middle height is $S=3.81 \pm 0.02$ mm, and 
        the width is $W=2.33 \pm 0.10$ mm.
    \item \textbf{Crescent moon-shaped objects:} 
        The length is $L=10.18 \pm 0.02$ mm, 
        the total height is $H=4.46 \pm 0.09$ mm, 
        the middle height is $S=3.91 \pm 0.14$ mm, and 
        the middle width is $W=2.13 \pm 0.02$ mm.
    \item \textbf{Open rings:} 
        We measured length $L$ and height $H$ of the rings as $L=5.45 \pm 0.01$ mm, $L=5.92 \pm 0.04$ mm, $L=6.48 \pm 0.08$ mm, $L=6.81 \pm 0.02$ mm, and $H=4.95 \pm 0.01$ mm, $H=4.90 \pm 0.02$ mm, $H=4.88 \pm 0.03$ mm, $H=4.86 \pm 0.02$ mm, respectively, for the rings with the openings of $1.02 \pm 0.05$ mm, $2.02 \pm 0.06$ mm, $3.02 \pm 0.05$ mm, and $4.02 \pm 0.05$ mm, respectively.  The width of all the rings remains constant at $W=0.596 \pm 0.005$ mm.
\end{enumerate}

The physical properties of the objects: mass, volume, and the total force $F$ equal to gravity minus buoyancy are presented in table~\ref{tab:phys_properties}. The masses of the objects were obtained using a high-performance analytical balance with a measuring accuracy of 0.1 mg per measurement. However, the masses of individual objects of the same type varied. Therefore, the mass measurements were conducted on at least ten objects of the same type, and the average values and their standard deviations were listed in table~\ref{tab:phys_properties}. The volume of the objects was measured experimentally using the water displacement method, a well-established technique for irregularly shaped solids. For the arrowhead and crescent moon,  we used 50 particles and assumed the displaced water volume to be equal to the total volume of all the particles.  Due to the unavailability of materials from the same batch, we used only 12 cone particles for volume measurement.  In this way, we determined the average volume of a single object. However, for the rings, the volume was calculated using the geometric formula of a torus. Please note that each object was used only once in the experiment, and we approximated its mass and volume by the average values, which were also used to evaluate the average total force $F$ (gravity minus buoyancy). 
\begin{table}[h!]
    \centering
    \begin{tabular}{c|c|c|c}
        Object (shape)  & Mass ($10^{-3}$ g) & Volume ($10^{-9}$ m$^3$) & Total force $F$ (mN) \\ 
        Cone             & 205.9 $\pm$ 1.3   & $97 \pm 20  $ & $1.10 \:\: \pm 0.19 \:\: $ \\
        Crescent moon    & 112.9 $\pm$ 0.3   & $55 \pm  7  $ & $0.59 \:\: \pm 0.07 \:\: $ \\
        Arrowhead        & 249.7 $\pm$ 0.8   & $50 \pm  7  $ & $1.98 \:\: \pm 0.07 \:\: $ \\
        Ring             & \: 30.6 $\pm$ 0.3 & $3.78 \pm 0.06  $ & $0.264 \pm 0.004 $
    \end{tabular}
    \caption{
    Physical properties of the various objects used during the experimental study. 
    }
    \label{tab:phys_properties}
\end{table}

\subsection{Methodology}
The experiments were performed by manually dropping the object from the top of the tank, approximately at the centre of the fluid surface.
The cameras were triggered when the object inside the tank was about to reach the visible zone of the cameras, and stopped when the object came out of the camera's image. 

Before performing each series of the experimental trials, we put the ruler at the central vertical line of the outside container walls facing both cameras, and we adjusted the camera positions to reach the same height $h$ of the recorded images from both cameras.  The motion of the objects took place close to the central vertical line of the container. The real height $h'$ of the image at this line was larger than $h$ owing to a finite camera viewing angle, decreased by the refraction of the light (the refractive index of the silicon oil is 1.4). For our setup, we measured the real image height $h'$ by putting the ruler inside the tank at its central line, and measuring the apparent image height $h$ by putting the other ruler at the central vertical line of the outside front wall of the container. In this way, we determined the correction factor $h'/h=1.077$, and we used this factor to recalculate the time-dependent particle positions from pixels to millimetres. 
Approximately one pixel was around 0.052 mm, but this calibration factor was determined specifically for each series of the experiments. 

The motion and reorientation of objects were recorded in the time sequence of the camera images. 
Further, all the images were transferred into MATLAB, and image analysis was performed to identify the objects there.  
Image processing steps involved the binarisation of the RGB images, background subtraction, and thresholding to obtain clearer shapes of the objects.
Additionally, noise removal techniques were employed to avoid spurious object detection in the cameras caused by shadows and reflections of the objects under high illumination.

\section{Experimental results}\label{eresults}

\subsection{Specification of the initial orientations}

In the preliminary trials, it was observed that all the shapes approached a stationary orientation for different initial configurations. Based on this feature, we conducted experimental trials for three distinct families of initial orientations at the fluid free surface\footnote{Note that the cameras started to record the particles after they settled around 10~cm from the fluid free surface.}, 
\begin{itemize}
    \item stationary, 
    \item inverted, 
    \item inclined. 
    \end{itemize}
    
The ``stationary'' initial orientation, with $x=X$, $y=Y$, and $z=Z$, was chosen to confirm that this orientation indeed remains stationary. We performed 27 such experiments, and indeed, in all of them, the particle orientation did not change with time.

The ``inverted'' initial orientation, with $x=-X$, $y=Y$, and $z=-Z$, was chosen aiming to observe the particle rotation around $y=Y$ axis, perpendicular to the plane of view of camera 1 (see figure~\ref{fig:sing_conv:mu51}, and therefore to measure the rotational-translational mobility coefficient $\mu_{51}$ from the theoretical relation \eqref{so51}. In this case, the centre of mass moves in the $xz$ plane, equal to the $XZ$ plane.

The ``inclined'' initial orientation, with $x=X$, and the angle $\theta$ between $z$ and $Z$ axes larger than $\pi/2$ and close to $\pi$, 
was chosen aiming to observe the particle rotation around $x=X$ axis, perpendicular to the plane of view of camera 2 (see figure~\ref{fig:sing_conv:mu42}, and therefore to measure the rotational-translational mobility coefficient $\mu_{42}$ from the theoretical relation \eqref{so42}. In this case, the centre of mass moves in the $yz$ plane, equal to the $YZ$ plane.

\subsection{Results for the stationary and inverted initial orientations}

We recorded the reorientation of sedimenting single cones, crescent moons, arrowheads, and open rings in 66 ``inverted'' experimental trials.
In all the trials performed for the same type of particle, it was observed that the particle always approached the same stationary orientation and remained there for a long time. However, in some cases, the particle motion was more complex than it had been predicted, involving also other components of the angular velocity, and a few trials could not be analysed quantitatively because the reorientation was only partially captured. For quantitative analysis of the motion of the cones, arrowheads, and open rings, we selected such trials for which the particle's rotation was only around the $y=Y$ axis. This choice allowed us to apply the simple evolution equation \eqref{so51} for the time-dependent inclination angle $\theta$ and determine the dimensionless rotational-translational mobility coefficient $\mu_{51}$. However, for the crescent moons, there were no such trials, even though we performed a large number of experiments. Instead, in some trials camera 1 recorded the rotation of the crescent moons only around $x=X$ axis, and we selected these experiments and compared the measured angle $\theta(\tilde{t})$ with equation~\eqref{so42}, extracting value of the other dimensionless rotational-translational mobility coefficient $\mu_{42}$. 

To illustrate how shapes reorient with time, for each trial we combined on a single image the sequences of both camera processed images, separated by the same time interval $\Delta \tilde{t}$ (equal to  1~s for arrowhead, 2~s for crescent moon, 4~s for cone and all the rings; the times of recording for each shape were: For the cone: 68~s, for the arrowhead: 40~s, for the crescent moon: 105~s, for the ring with 1~mm opening: 180~s, for the ring with 2~mm opening: 188~s, for the ring with 3~mm opening: 190~s, for the ring with 4~mm opening: 196~s). The superposition was made using the MATLAB library's {\tt imfuse} function, keeping the same scale and size of all the images. The resulting time sequences of the snapshots of the sedimenting particle for all the trials are shown in the repository \cite{repository}. In this paper, we present a few of them, after inverting black and white, and with a trimmed width, keeping only $\pm 3$~cm from the central vertical line of each image. 

Examples of snapshots from the ``selected inverted'' trials showing the reorientation behaviour for various shapes are presented in figures~\ref{fig:sedim_CMA} and \ref{fig:sedim_rings}. 
In Appendix~\ref{sec:other_behaviour}, we present two examples of the 3-dimensional dynamics, starting from the inverted initial orientation, with non-vanishing and time-varying three components of the angular velocity.
The overall statistics of all the trials can be found in Appendix~\ref{number}. 
\begin{figure}
    \centering
    \begin{subfigure}{\textwidth}
        \centering
        \includegraphics[clip, trim=2.1cm 2.1cm 2.1cm 2.1cm, width=0.9\linewidth]{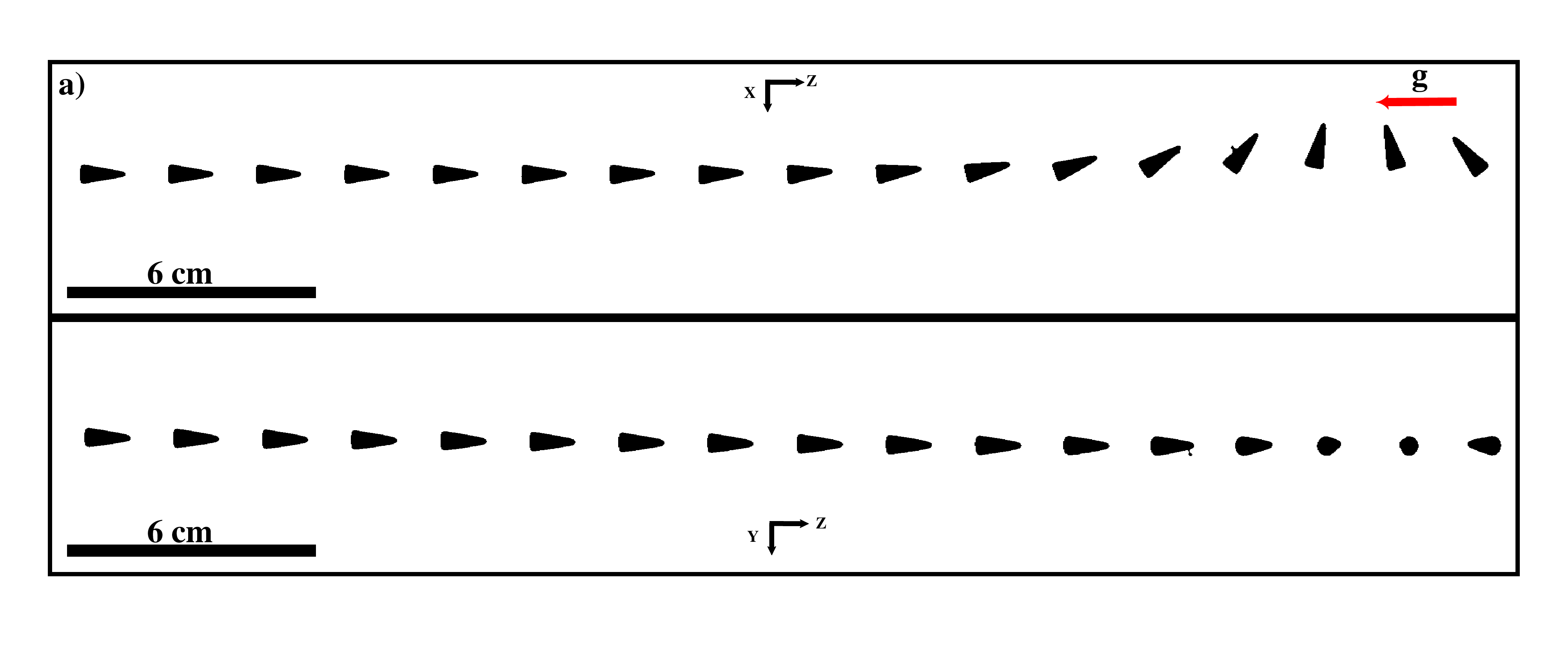}
        \phantomsubcaption
        \label{fig:sedim_CMA:a}
    \end{subfigure}
    \begin{subfigure}{\textwidth}
        \centering
        \includegraphics[clip, trim=2.1cm 2.1cm 2.1cm 2.1cm, width=0.9\linewidth]{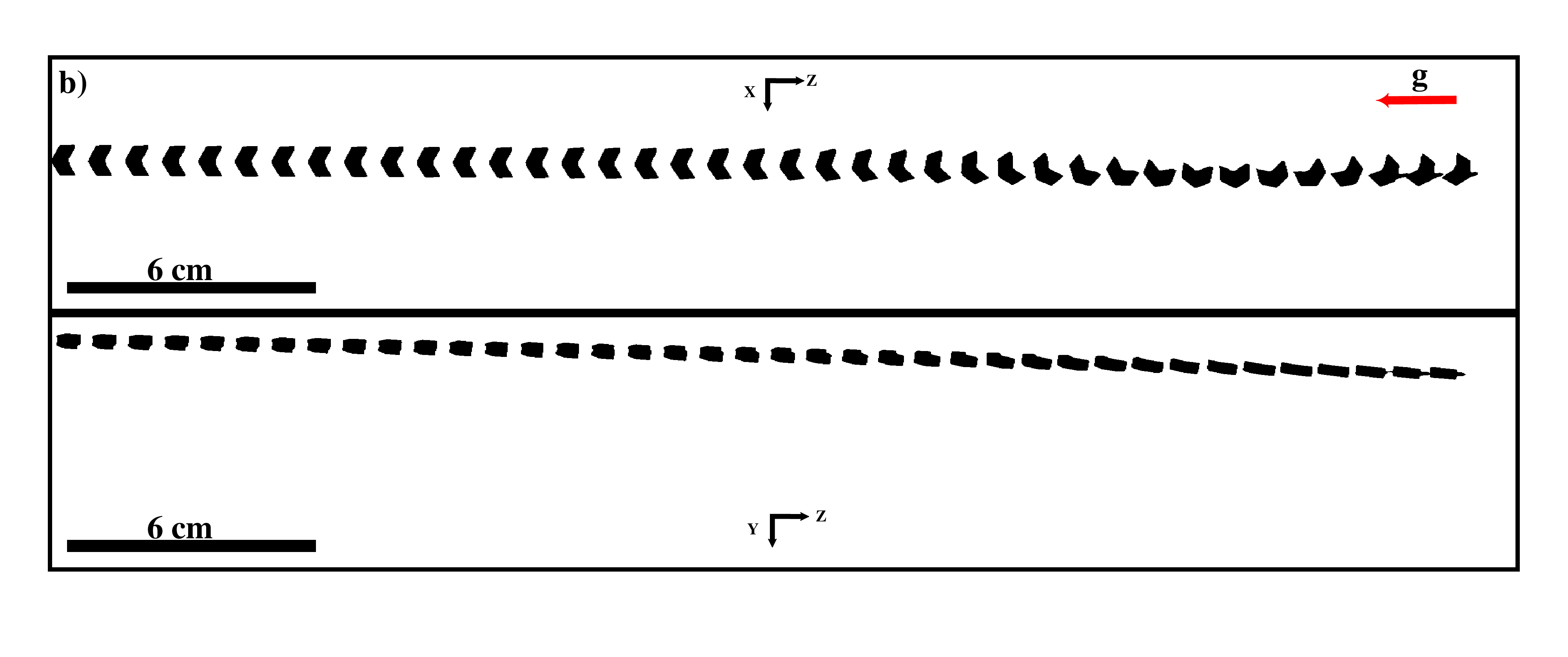}
        \phantomsubcaption
        \label{fig:sedim_CMA:b}
    \end{subfigure}
    \begin{subfigure}{\textwidth}
        \centering
        \includegraphics[clip, trim=2.1cm 2.1cm 2.1cm 2.1cm, width=0.9\linewidth]{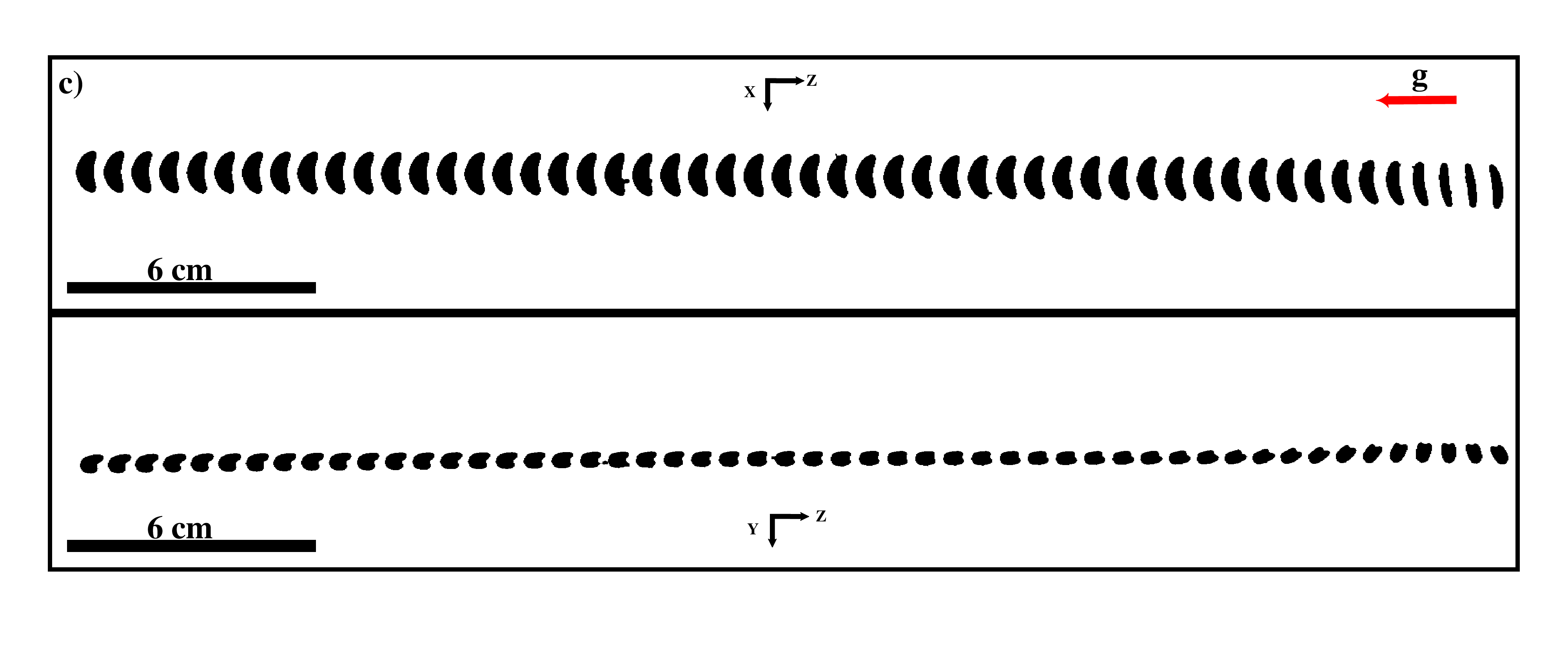}
        \phantomsubcaption
        \label{fig:sedim_CMA:c}
    \end{subfigure}
    \caption{
        Snapshots from 3 experimental trials showing time-dependent position and orientation of a single sedimenting object: a) cone, b) arrowhead, and c) crescent moon. 
        The images were recorded simultaneously by two cameras (top and bottom panels). 
        Gravity points left, and the particles move from right to left.
    }
    \label{fig:sedim_CMA}
\end{figure}
\begin{figure}
    \centering
    \begin{subfigure}{\textwidth}
        \centering
        \includegraphics[clip, trim=2.1cm 2.1cm 2.1cm 2.1cm, width=0.9\linewidth]{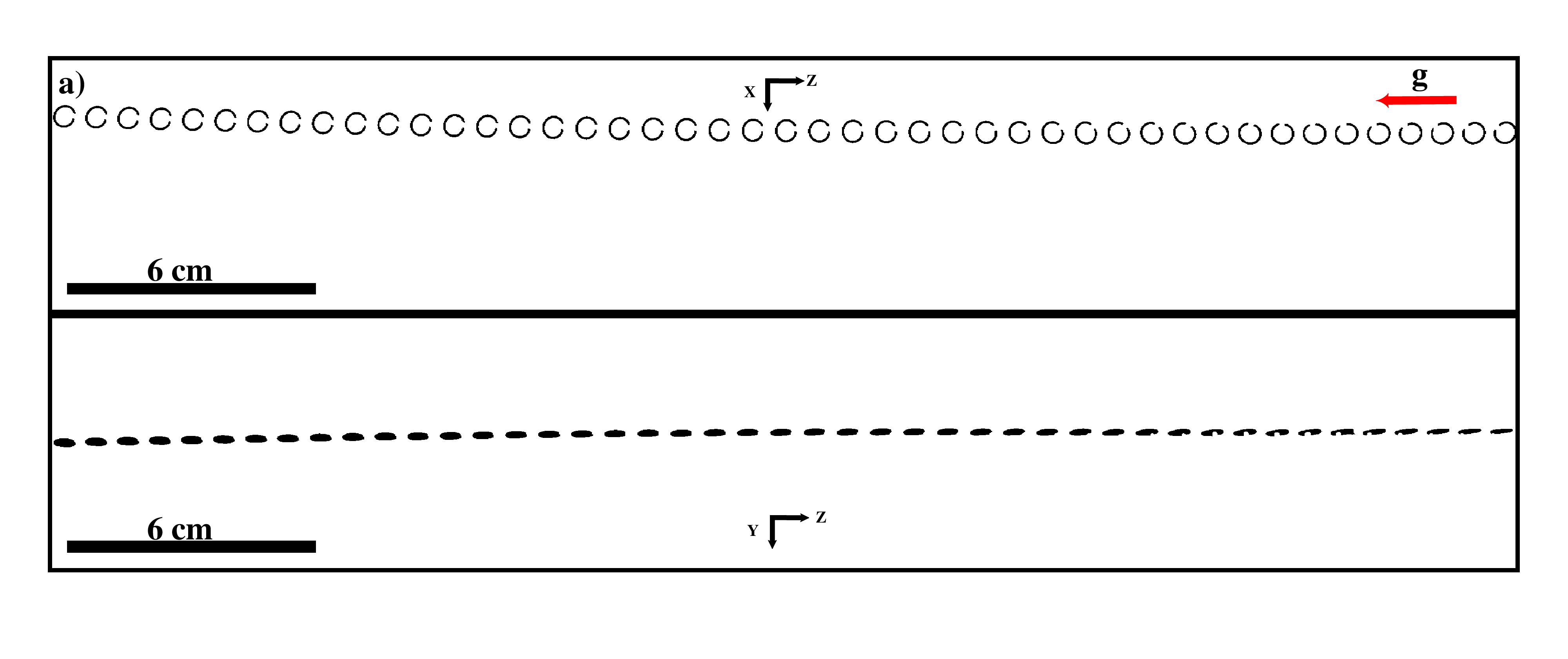}
        \phantomsubcaption
        \label{fig:sedim_rings:a}
    \end{subfigure}
    \begin{subfigure}{\textwidth}
        \centering
        \includegraphics[clip, trim=2.1cm 2.1cm 2.1cm 2.1cm, width=0.9\linewidth]{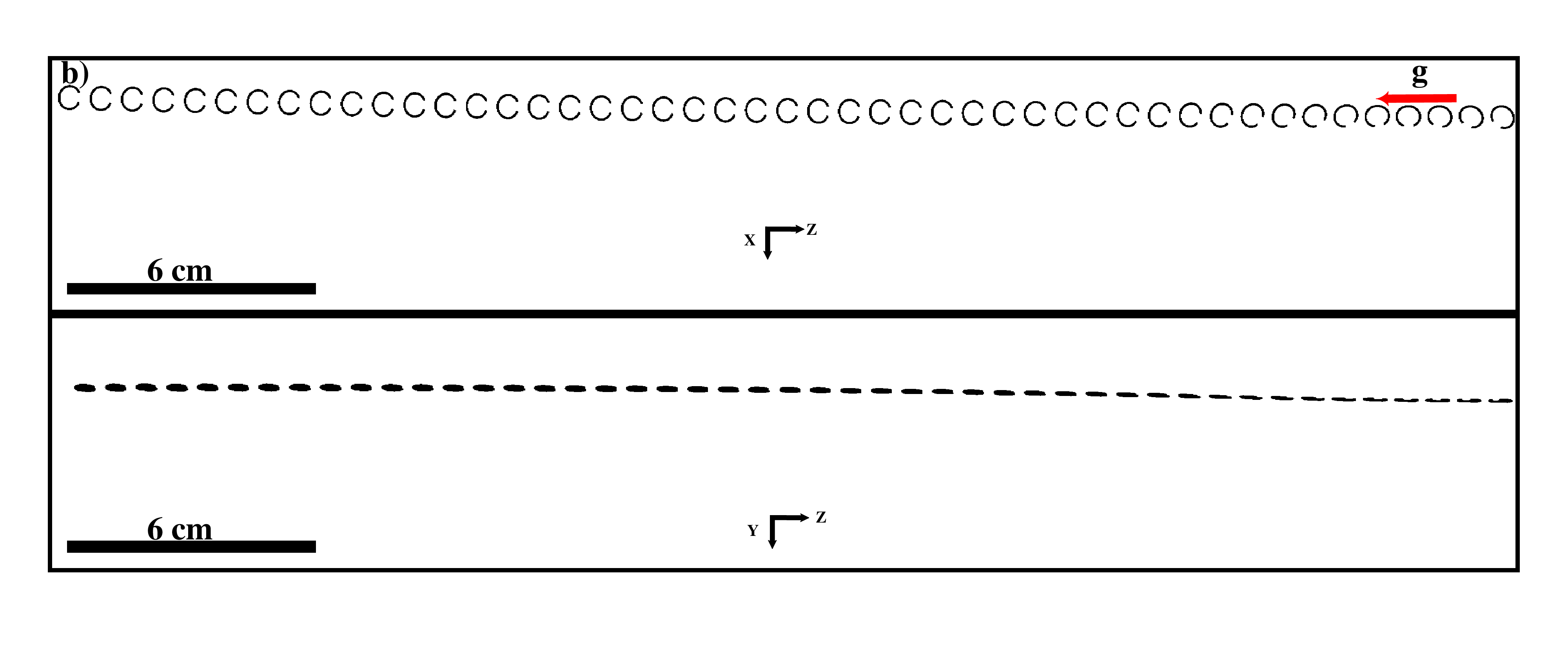}
        \phantomsubcaption
        \label{fig:sedim_rings:b}
    \end{subfigure}
    \begin{subfigure}{\textwidth}
        \centering
        \includegraphics[clip, trim=2.1cm 2.1cm 2.1cm 2.1cm, width=0.9\linewidth]{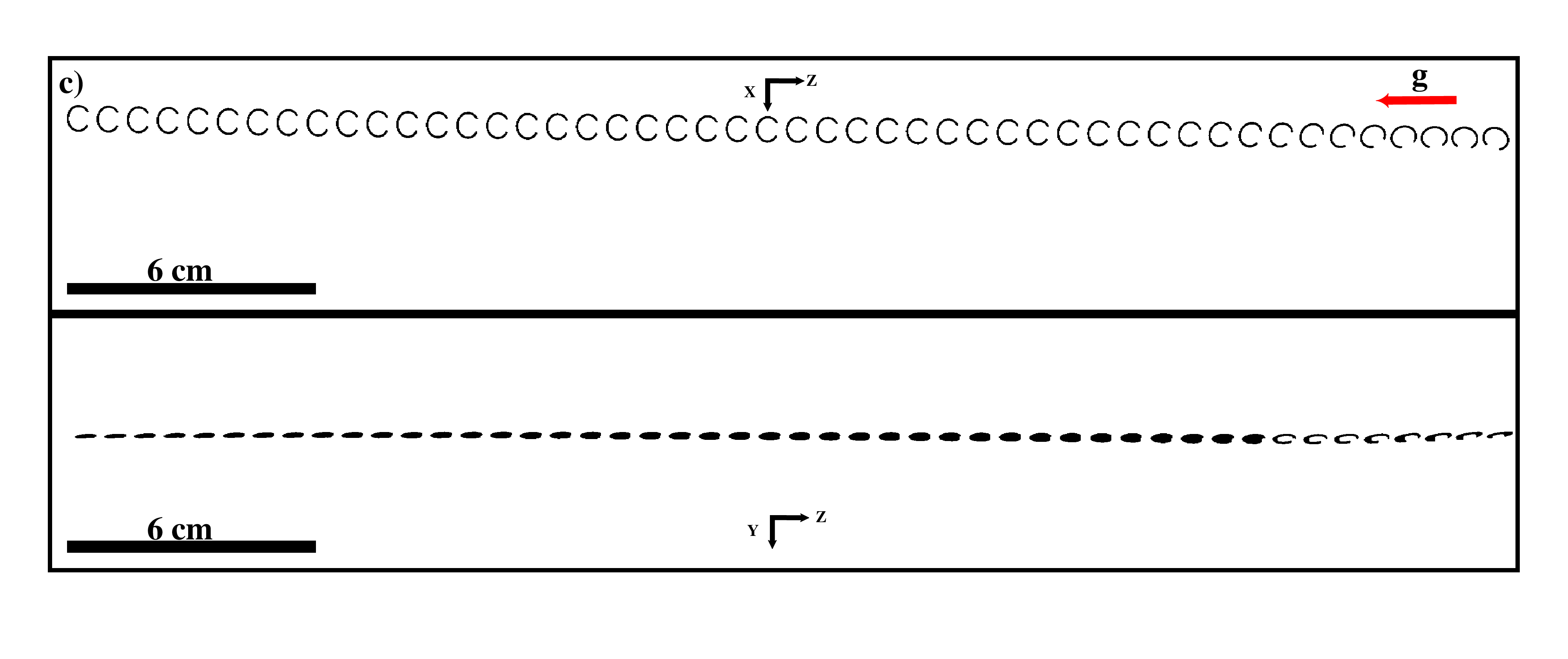}
        \phantomsubcaption
        \label{fig:sedim_rings:c}
    \end{subfigure}
    \begin{subfigure}{\textwidth}
        \centering
        \includegraphics[clip, trim=2.1cm 2.1cm 2.1cm 2.1cm, width=0.9\linewidth]{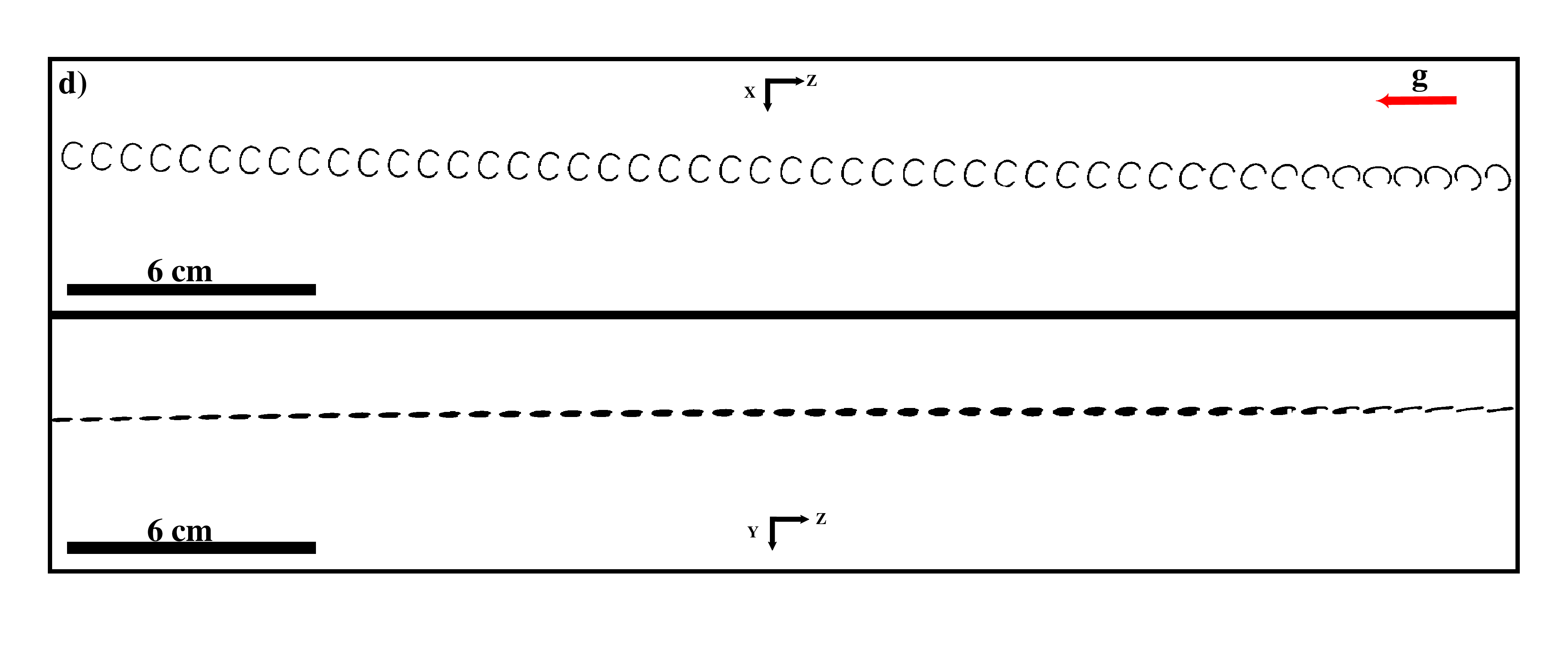}
        \phantomsubcaption
        \label{fig:sedim_rings:d}
    \end{subfigure}
    \caption{
        Snapshots from 4 experimental trials showing time-dependent position and orientation of a single sedimenting ring with the opening of: a) 1 mm, b) 2 mm, c) 3 mm, d) 4 mm. 
        The images were recorded simultaneously by camera 1 (top panel) and camera 2 (bottom panel). Gravity points left, and the particles move from right to left.
    }
    \label{fig:sedim_rings}
\end{figure}

For the selected trials, following the accurate identification of object shapes across all image sequences, a quantitative analysis of all the images from camera 1 was performed. 
First, a few selected geometric properties of the objects were identified at each time instance $\tilde{t}$ using MATLAB's built-in function {\tt regionprops}, including the object's centroid position in pixels, next converted via the calibration factor to millimetres resulting in $(X_c, Z_c)$, and object's inclination angle $\theta$, as explained in Appendix~\ref{app:extracting:1}.
The time-dependent vertical velocity component $U_Z(\tilde{t})$ was calculated as the difference $Z_c(\tilde{t}+\Delta \tilde{t})-Z_c(\tilde{t})$, divided by the time interval between the consecutive images, i.e., by $\Delta \tilde{t}=1$~s. 
The orientation angle of an object is one of the direct outputs of MATLAB's built-in function {\tt regionprops}, which uses the covariance matrix $\boldsymbol{\Sigma}$ constructed from the normalised central moments of the second order obtained from the binary image, as described in Appendix~\ref{app:extracting:1}.
In addition to the orientation angle, its uncertainty is determined using a custom function that analyses the pixel distribution of the object.
This uncertainty is a function of the eigenvalues of the covariance matrix $\boldsymbol{\Sigma}$ and the total number of pixels (all are outputs of {\tt regionprops}), as derived in Appendices~\ref{app:extracting:2} and~\ref{app:extracting:3}.
Finally, the uncertainty for the values of the function $\tan(\theta/2)$ that will be used later is computed using the standard propagation of error method as shown in Appendix~\ref{app:extracting:3}.
\begin{figure}[h!]
    \centering
    \includegraphics[clip, trim=5.1cm 0.8cm 5cm 1.2cm, width=0.96\linewidth]{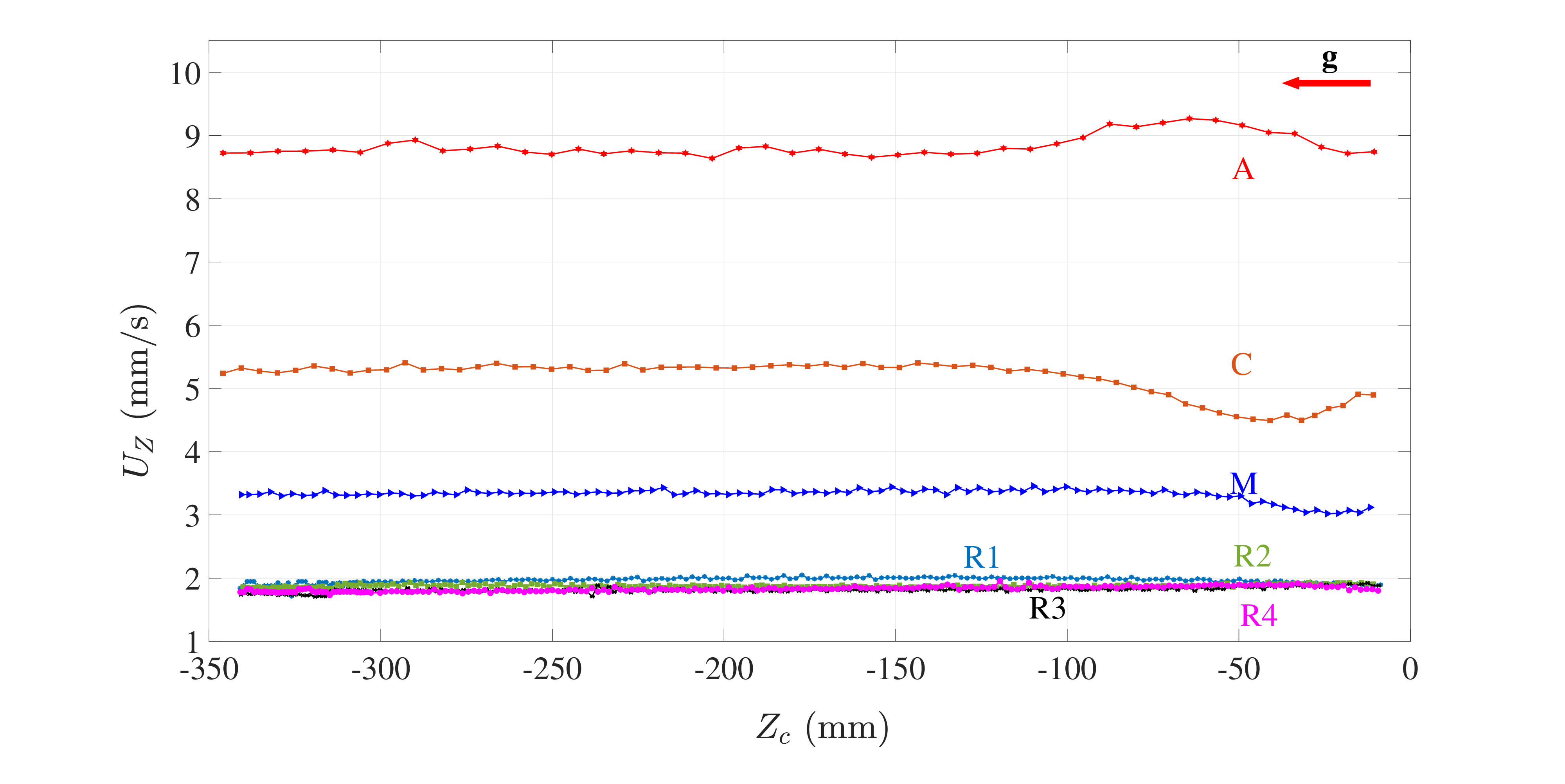} 
    \vspace{-0.3cm}
    \caption{
    Vertical component $U_Z$ of the sedimentation velocity of the object's centre of mass for the experimental trials shown in figures~\ref{fig:sedim_CMA}-\ref{fig:sedim_rings}. 
    Names of the objects: C - cone, A - arrowhead, M - crescent moon, R$i$ - ring with the opening width equal to $i = 1, 2, 3, 4$ mm.  
    } 
    \label{velocity_vs_t}
\end{figure}

Figure \ref{velocity_vs_t} presents the vertical component $U_Z$ of the object's centre velocity as a function of the object's vertical position $Z_c$ for the experimental trials shown in figures~\ref{fig:sedim_CMA}-\ref{fig:sedim_rings}. 
Negative values indicate a downward measurement direction (from top to bottom of the tank). Comparing figure~\ref{velocity_vs_t} with the snapshot sequences in figures~\ref{fig:sedim_CMA}-\ref{fig:sedim_rings}, we observe that for the cone and the crescent moon $U_Z$  decreases and reaches a minimum when the cone and the crescent moon orients horizontally at $\theta = \pi/2$, and then increases when $\theta \rightarrow 0$. On the contrary, for the arrowhead $U_Z$ increases and reaches a maximum when the particle orients vertically at $\theta = \pi/2$, and then decreases when $\theta \rightarrow 0$. This is in agreement with the well-known rule that elongated particles oriented vertically sediment faster than if oriented horizontally. 

\begin{figure}
    \centering
    \includegraphics[clip, trim=8cm 0cm 10cm 1cm, width=0.9\linewidth]{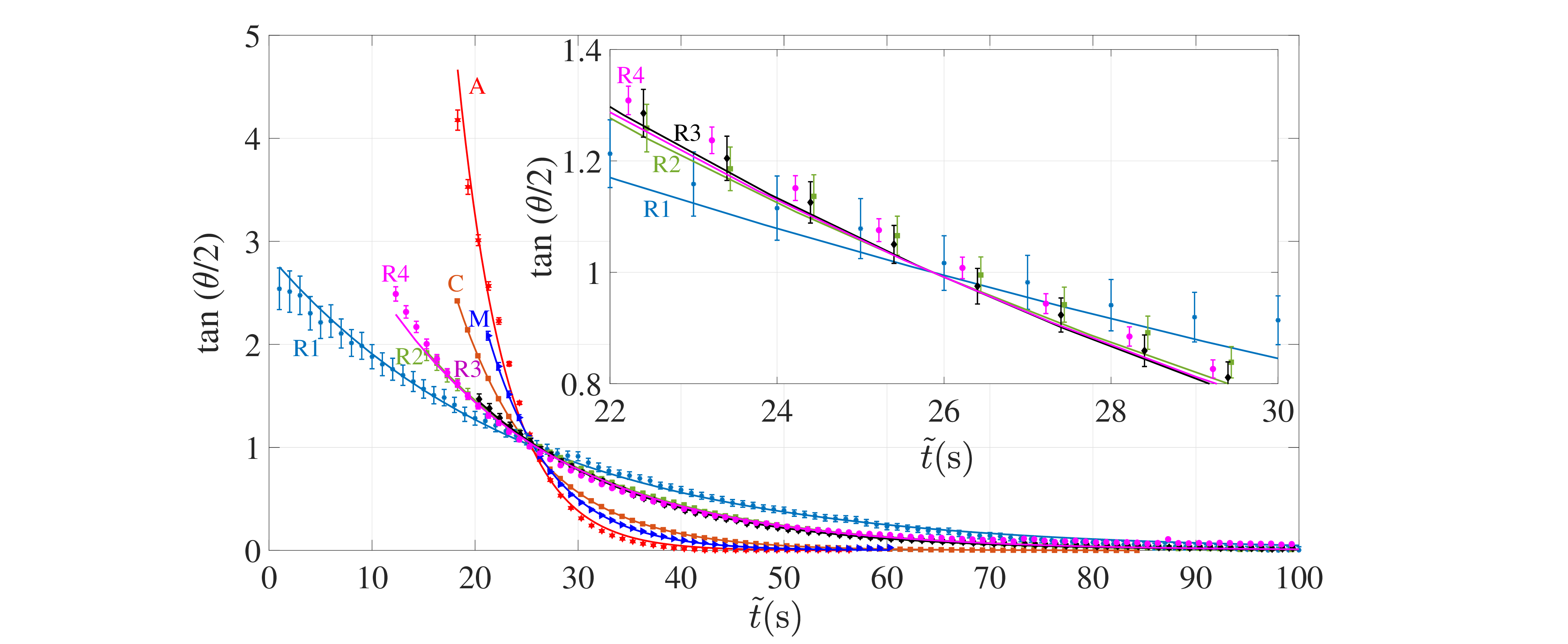} 
    \vspace{-0.3cm}
    \caption{
    Reorientation of particles of different shapes in the experimental trials shown in figures~\ref{fig:sedim_CMA} and \ref{fig:sedim_rings}. 
    The time dependence of $\tan(\theta/2)$ for the experimental data (symbols) is approximated (solid lines) by the exponential function \eqref{eq:exp_exp} with the fitted values of the parameters $A$ and $\tilde{\tau}$, listed in Appendix~\ref{sec:experimental_parameters}. 
    Names of the objects: C - cone, A - arrowhead, M - crescent moon, R$i$ - ring with the opening width equal to $i = 1, 2, 3, 4$ mm. 
    The plots for different objects are shifted in time to reach $\tan(\theta/2)=1$ at the same time instant.
    }
    \label{tan_theta_vs_t}
\end{figure}

The reorientation of the particles over time is quantified, as shown in figure~\ref{tan_theta_vs_t}. 
Here, we plot $\tan(\theta/2)$ as a function of dimensional time $\tilde{t}$ for the trials shown and analysed in figures~\ref{fig:sedim_CMA}-\ref{velocity_vs_t}.
Note that the uncertainty for the values of the function $\tan(\theta/2)$ is computed using the standard propagation of error method as shown in Appendix~\ref{app:extracting:3}.
For a better comparison, we shifted the plots of different objects in time so that they reach the value $\tan(\theta/2)=1$ at the same instant. 
To determine the rotational-translational mobility coefficients, we use the theoretically derived time-dependence \eqref{so51}-\eqref{so42} of $\tan(\theta/2)$. 
Therefore, we fit the experimental data with the exponential decay function:
\bee
    && \tan(\theta/2)= A \exp( - \tilde{t} / \tilde{\tau} ),
    \label{eq:exp_exp}
\eee
where the amplitude $A$ and the characteristic time $\tilde{\tau}$ are determined using MATLAB's nonlinear data-fitting function {\tt fittype}.  
Depending on the particle, the inverse of the characteristic time $1/\tilde{\tau}$ is either the coefficient $1/|\tilde{\tau}_{51}|$ or $1/|\tilde{\tau}_{42}|$, see equation~\eqref{eq:mobility}.  
The fitting parameters, which correspond to the solid lines in figure~\ref{tan_theta_vs_t}, can be found in Appendix~\ref{sec:experimental_parameters}. Characteristic reorientation times depend on shape. They are the smallest for the arrowhead, larger for the crescent moon, still larger for the cone, and even larger for the rings, being the largest for the ring with the smallest opening 1~mm.

\subsection{Results for the inclined initial orientation}

\begin{figure}
    \centering
    \begin{subfigure}{\textwidth}
        \centering
        \includegraphics[clip, trim=2.1cm 2.1cm 2.1cm 2.1cm, width=0.9\linewidth]{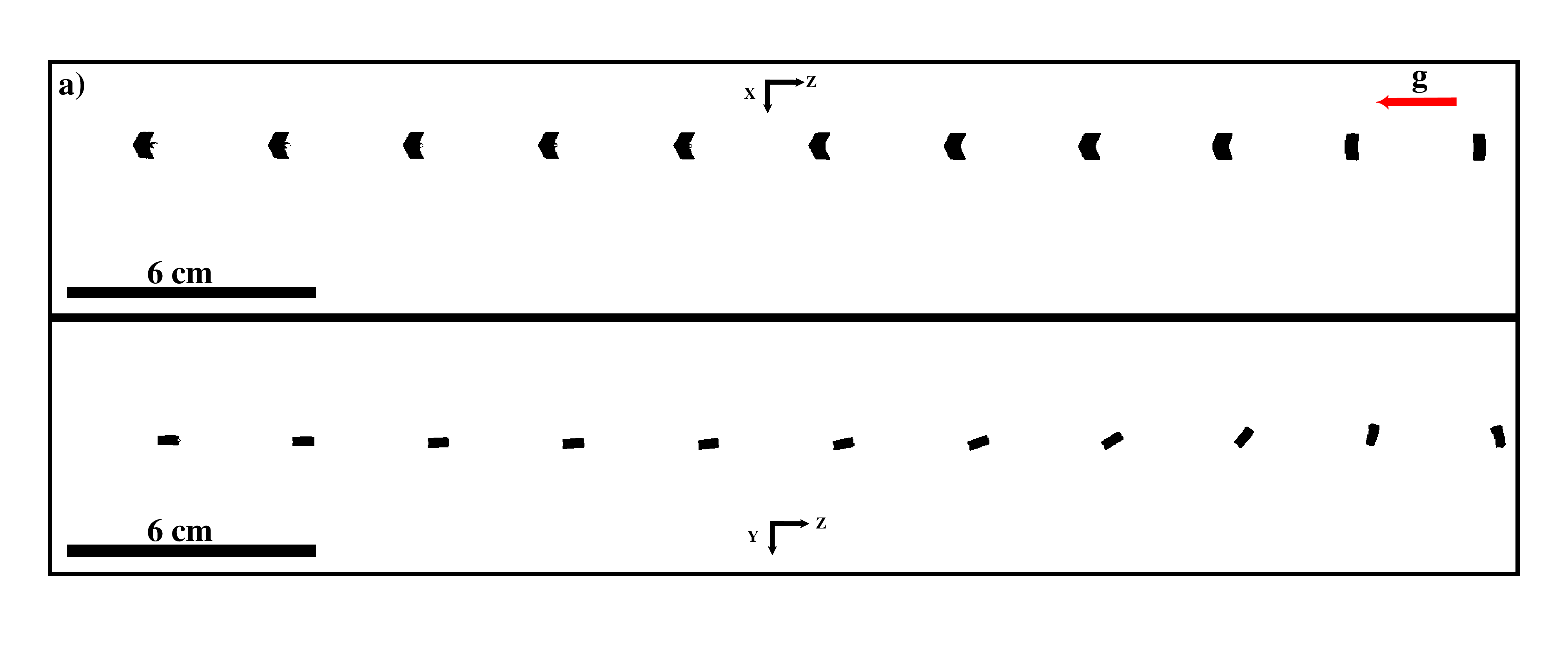} 
        \phantomsubcaption
        \label{fig:sedim_in_objects:a}
    \end{subfigure}
    \begin{subfigure}{\textwidth}
        \centering
        \includegraphics[clip, trim=2.1cm 2.1cm 2.1cm 2.1cm, width=0.9\linewidth]{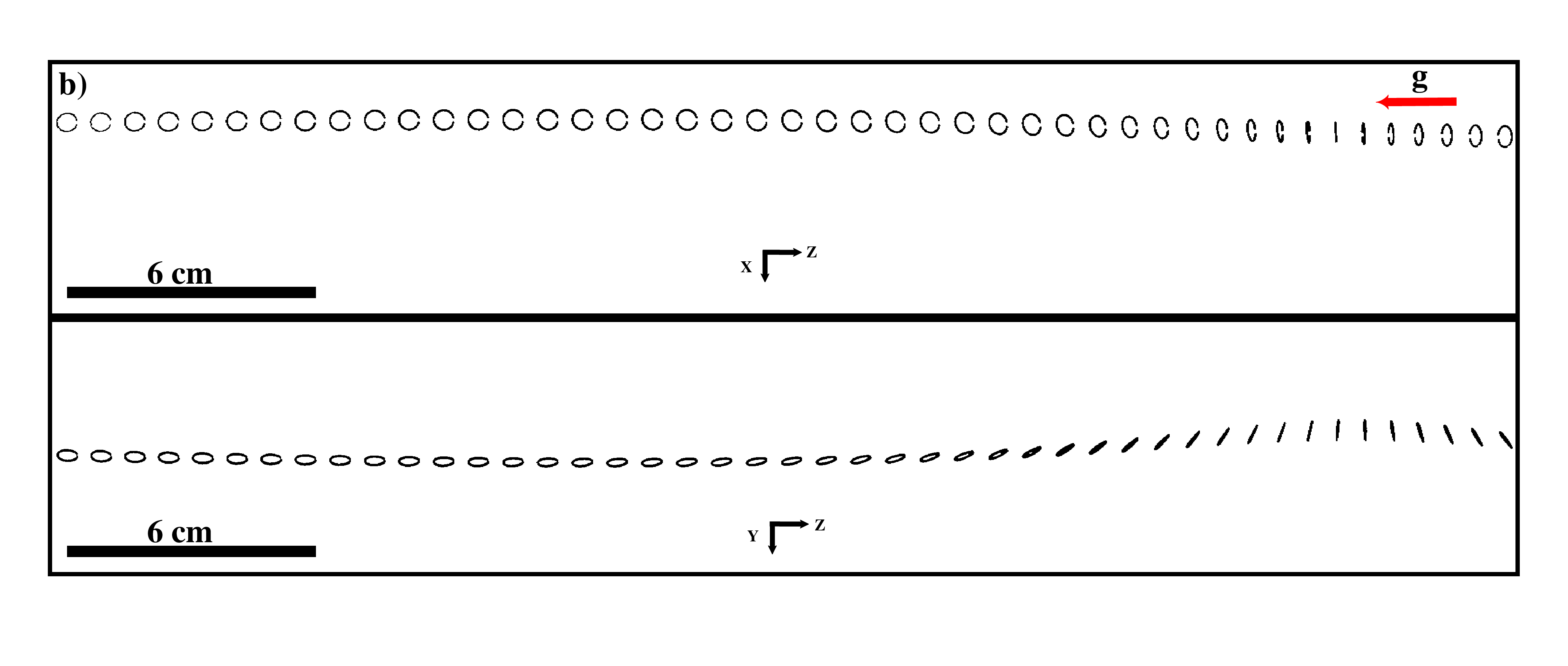}
        \phantomsubcaption
        \label{fig:sedim_in_objects:b}
    \end{subfigure}
    \caption{
        Snapshots from 2 experimental trials with the inclined initial orientation, showing time-dependent position and orientation of a single sedimenting object: a) arrowhead, b) ring with the opening of 1  mm. 
        The images were recorded simultaneously by two cameras (top and bottom panels).  
        Gravity points left, and the particles move from right to left.
    }
    \label{fig:sedim_in_objects}
\end{figure}

We performed 45 experimental trials for the rings and arrowheads, initially at the inclined orientation. 
For the quantitative analysis, we selected the trials in which the particle rotated around the $x=X$ axis. 
The snapshots from two selected trials are shown in figure~\ref{fig:sedim_in_objects}. 
We used equation~\eqref{so42} to determine the rotational-translational mobility coefficient $\mu_{42}$, proceeding in the same way as described for the inverted trials. 

In Appendix~\ref{sec:other_behaviour}, we present an example of the 3-dimensional dynamics, starting from the inclined initial orientation, with non-vanishing and time-varying three components of the angular velocity.
The overall statistics of all the trials can be found in Appendix~\ref{number}.

\subsection{Averaged sedimentation velocity and mobility coefficients}

To extract average values of the parameters of each object type, we selected from the inverted class of the initial orientations the experimental trials, for which the particle rotates only around the axis $x=X$ or only around the axis $y=Y$, and from the inclined class of the initial orientations, the trials for which the particle rotates only around the axis $x=X$ (see figure~\ref{fig:sing_conv:mu51}-\ref{fig:sing_conv:mu42}). 
In the following, we will call these trials selected inverted and selected inclined. 
To calculate the average velocity, we took into account the ``selected inverted'' and ``stationary'' trials (for the numbers of the trials, see Appendix~\ref{number}).

Table~\ref{tab:mobility_velocity} presents the average parameters. 
In the case of the terminal vertical velocity $U_{f}$ identified for a particular particle type, the procedure was the following. 
First, the time-average of vertical velocity $U_Z(\tilde{t})$ was computed for each ``stationary'' and each ``selected inverted'' trial. 
For a ``stationary'' trial, $U_Z(\tilde{t})$ was averaged over the whole range of time $\tilde{t}$. 
For a ``selected inverted'' trial, $U_Z({\tilde t})$ was averaged over time $\tilde{t}\ge \tilde{t}_o$,  only after the object had reached a stable, stationary state. 
In these cases, the time $\tilde{t}_o$ at which this occurred was identified manually for each trial.
The mean vertical terminal velocity $U_f$ and its standard deviation were then computed from the time-averaged vertical terminal velocities for all the selected trials.
The non-dimensional values of $\mu_{33}$ were then evaluated from equation~\eqref{eq:mju33}, with the uncertainties computed using the propagation of error approach.

\begin{table}
    \centering
    \begin{tabular}{c|c|c|c|c|c|c}
      Object & \multicolumn{3}{c|}{Dimensional form} & \multicolumn{3}{c}{Non-Dimensional form} \\  
      & $U_{\!f}$ (mm/s) 
      &  $-1/\tilde{\tau}_{51}$ (s$^{-1}$) 
      & $ 1/\tilde{\tau}_{42}$ (s$^{-1}$)  
      & ${\mu_{33}}$ 
      & ${-\mu_{51}}$ 
      & ${\mu_{42}}$ \\
     C & 5.23 $\pm$ 0.10 & 0.113 $\pm$ 0.009 & 0.113 $\pm$ 0.009 & 0.320 $\pm$ 0.055 & 0.030 $\pm$ 0.006 & 0.030 $\pm$ 0.006 \\
     M & 3.26 $\pm$ 0.19 & - & 0.144 $\pm$ 0.032 & 0.864 $\pm$ 0.111 & - & 0.389 $\pm$ 0.098 \\
     A & 8.72 $\pm$ 0.46 & 0.189 $\pm$ 0.016 & 0.129 $\pm$ 0.017 & 0.422 $\pm$ 0.027 & 0.057 $\pm$ 0.008 & 0.039 $\pm$ 0.005 \\
     R1 & 1.91 $\pm$ 0.07 & 0.042 $\pm$ 0.005 & 0.038 $\pm$ 0.006 & 0.598 $\pm$ 0.023 & 0.072 $\pm$ 0.008 & 0.065 $\pm$ 0.011\\
     R2 & 1.88 $\pm$ 0.03 & 0.055 $\pm$ 0.005 & 0.056 $\pm$ 0.02 & 0.642 $\pm$ 0.015 & 0.110 $\pm$ 0.010 & 0.114 $\pm$ 0.041\\
     R3 & 1.84 $\pm$ 0.05 & 0.059 $\pm$ 0.006 & 0.066 $\pm$ 0.009 & 0.685 $\pm$ 0.022 & 0.143 $\pm$ 0.014 & 0.159 $\pm$ 0.023\\
     R4 & 1.74 $\pm$ 0.05 & 0.066 $\pm$ 0.006 & 0.070 $\pm$ 0.008 & 0.682 $\pm$ 0.024 & 0.176 $\pm$ 0.016 & 0.186 $\pm$ 0.022 \\
    \end{tabular}
    \caption{
        The average parameters: terminal vertical velocity $U_{\!f}$ of the particle in the stationary configuration and coefficients $1/\tilde{\tau}_{51}$ or $1/\tilde{\tau}_{42}$ for each object type, together with the dimensionless mobility coefficients. 
        Names of the objects: C - cone, A - arrowhead, M - crescent moon, R$i$ - ring with the opening width equal to $i = 1, 2, 3, 4$ mm.
        }
    \label{tab:mobility_velocity}
\end{table}

The dimensional coefficients $1/\tilde{\tau}_{51}$ and $1/\tilde{\tau}_{42}$, determined for each selected inverted or selected inclined trial by fitting the time dependence of $\tan (\theta/2)$ with equation~\eqref{eq:exp_exp}, were later averaged over all the selected trials, separately for each type of the particle, with the mean values and their standard deviations reported in Table~\ref{tab:mobility_velocity}.
The non-dimensional values of the rotational-translational mobility coefficients $\mu_{42}$ and $\mu_{51}$ were then computed from equation~\eqref{eq:mobility}, with the uncertainties evaluated using the propagation of error approach.

\section{Theoretical modeling}\label{tresults}
Each of the rigid particles used in the experiments, shown in figure~\ref{objects}, is approximated as a large number of identical touching spherical beads of diameter $d$. 
The bead models of the particles are shown in figures~\ref{shapes_beads} and \ref{shapes_beads_moon}. 
The construction details are outlined in Appendix~\ref{sec:theor_param}. 
The generalisation of the model shown in figure~\ref{fig:Ring1mm}-\ref{fig:Ring3mm} for the rings with the openings of 2~mm and 4~mm is straightforward.
\begin{figure}[h!]
    \centering
    \begin{subfigure}{0.37\textwidth}
        \centering
        \includegraphics[width=\textwidth]{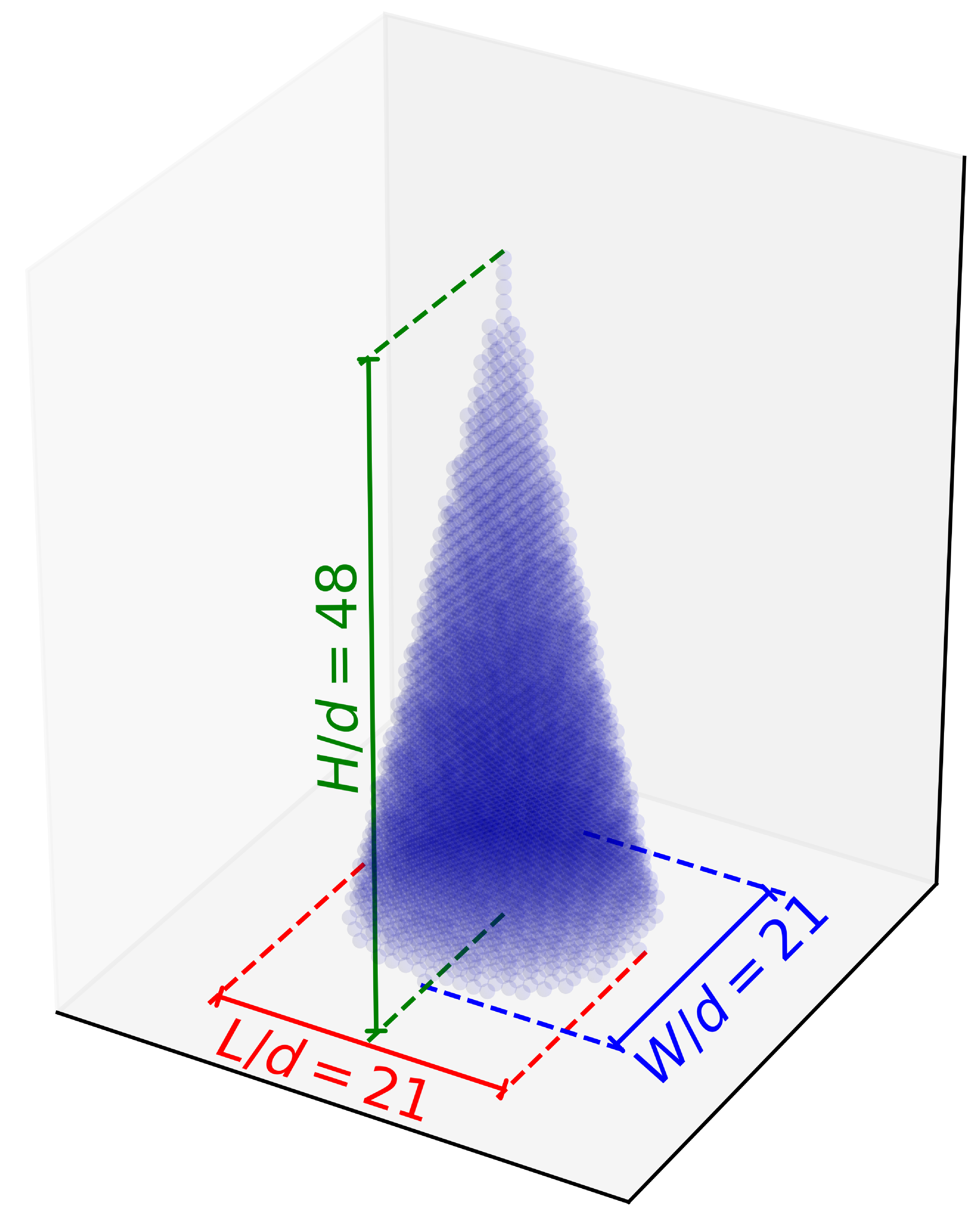}
        \caption{Cone: $N=5088$}
        \label{fig:Cone}
    \end{subfigure}
    \hfill
    \begin{subfigure}{0.45\textwidth}
        \centering
        \includegraphics[width=\textwidth]{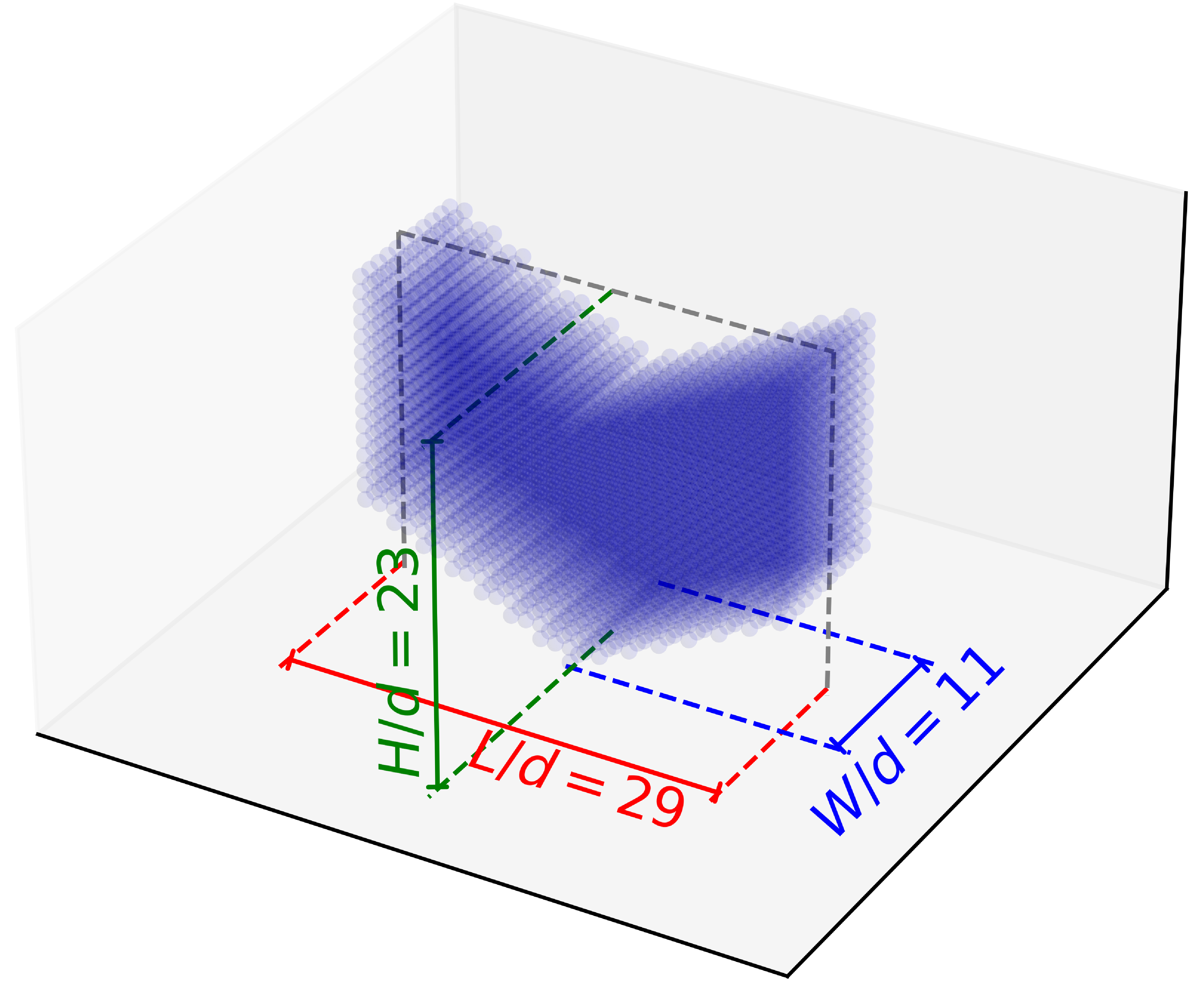}
        \caption{Arrowhead: $N=5104$}
        \label{fig:Arrow}
    \end{subfigure}
        \begin{subfigure}{0.41\textwidth}
        \centering
        \includegraphics[width=\textwidth]{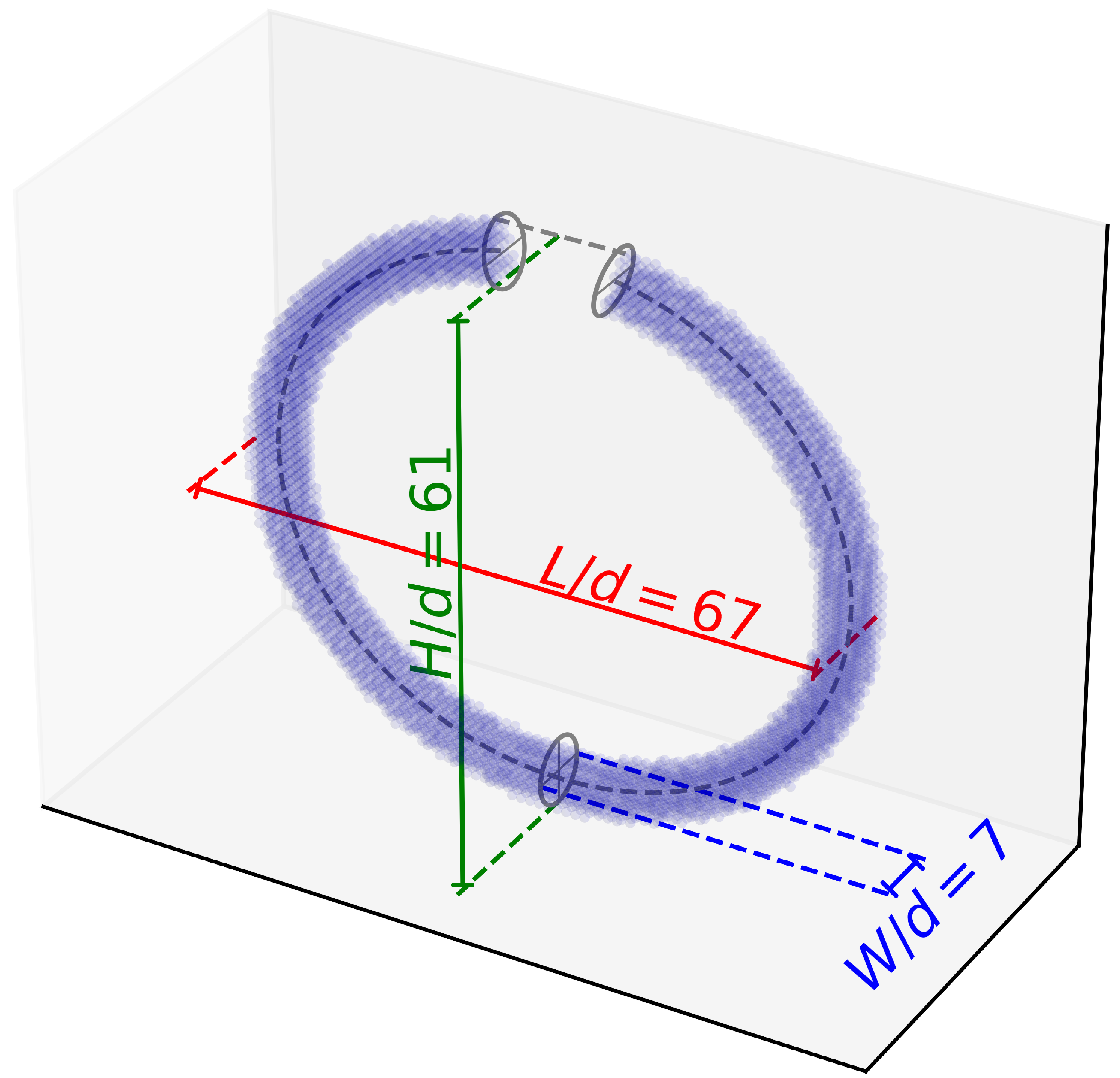}
        \caption{Open ring with a 1~mm-wide gap: $N=5182$}
        \label{fig:Ring1mm}
    \end{subfigure}\hspace{2.5cm}
    \begin{subfigure}{0.41\textwidth}
        \centering
        \includegraphics[width=\textwidth]{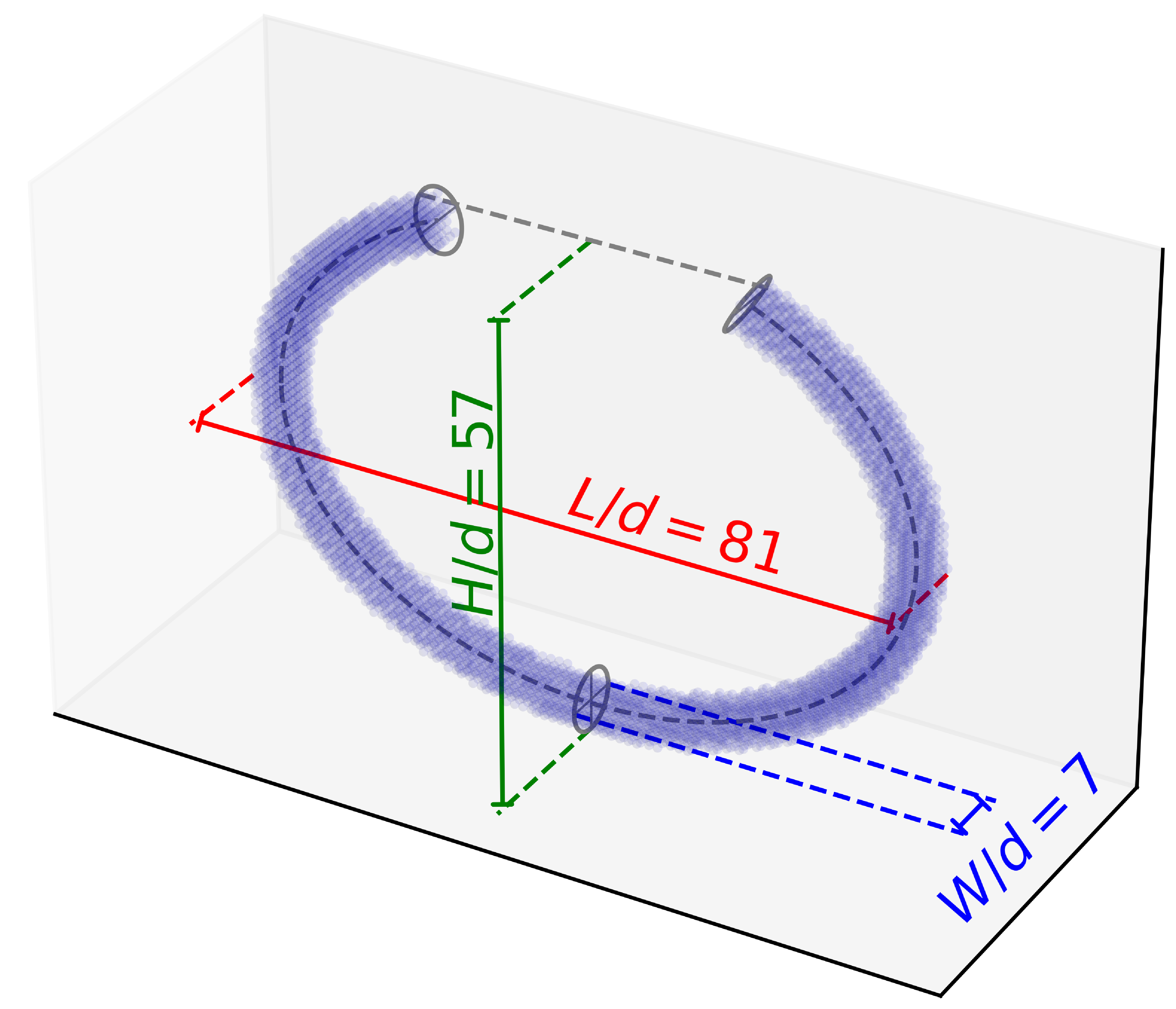}
        \caption{Open ring with a 3~mm-wide gap: $N=5110$}
        \label{fig:Ring3mm}
    \end{subfigure}
    \caption{
    The bead model of the experimental rigid particles (for the crescent moon, see figure~\ref{shapes_beads_moon}). The particles are shown in their stationary stable configuration at $\theta=0$, with $H$ along gravity.
    }
    \label{shapes_beads}
\end{figure}

\begin{figure}[h!]
    \centering
    \begin{subfigure}{\textwidth}
        \centering
        \includegraphics[width=0.55\textwidth]{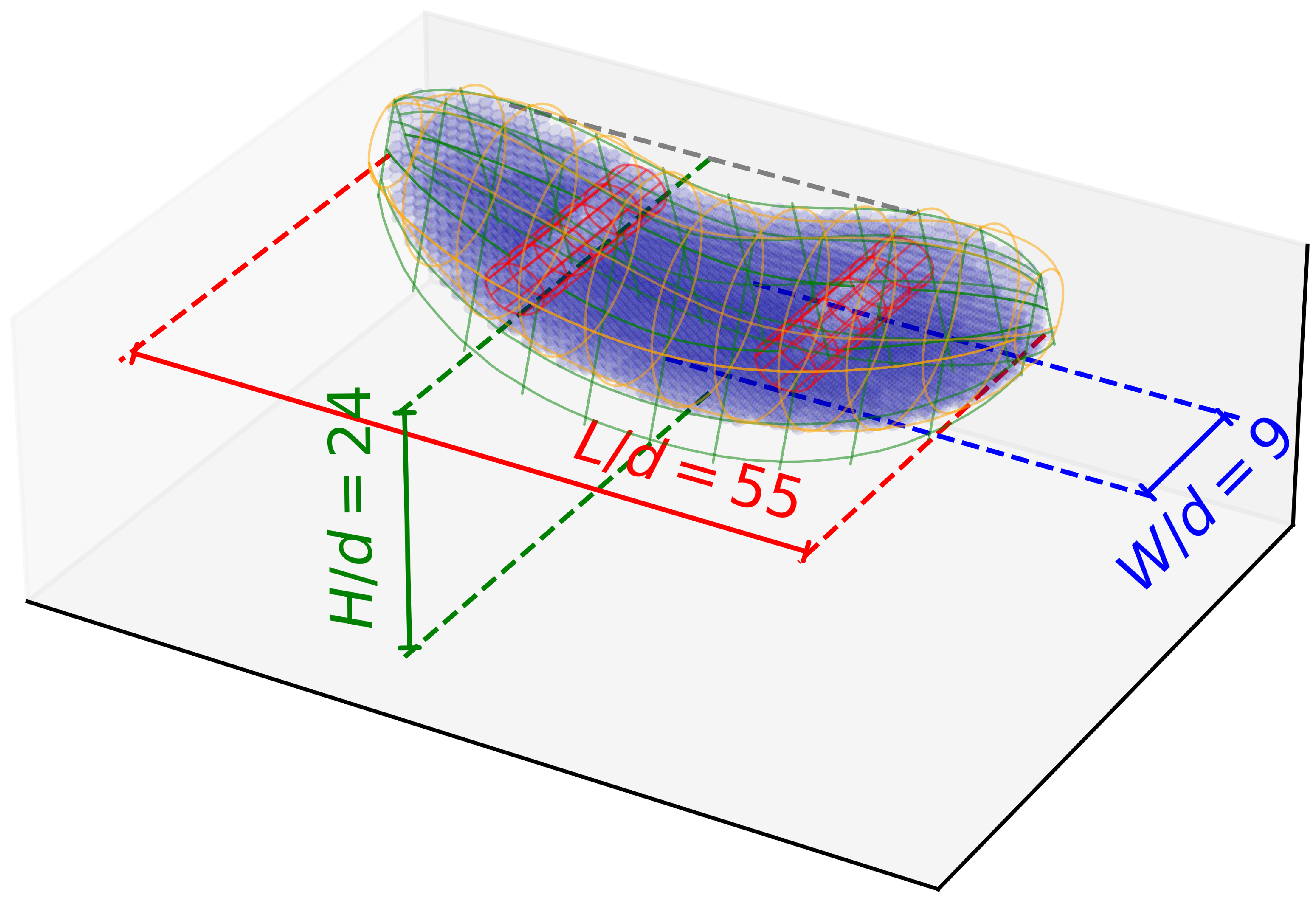}
        \caption{Crescent moon: $N=5513$}
        \label{fig:Moon}
    \end{subfigure}
    \begin{subfigure}{\textwidth}
        \centering
        \includegraphics[width=0.7\textwidth]{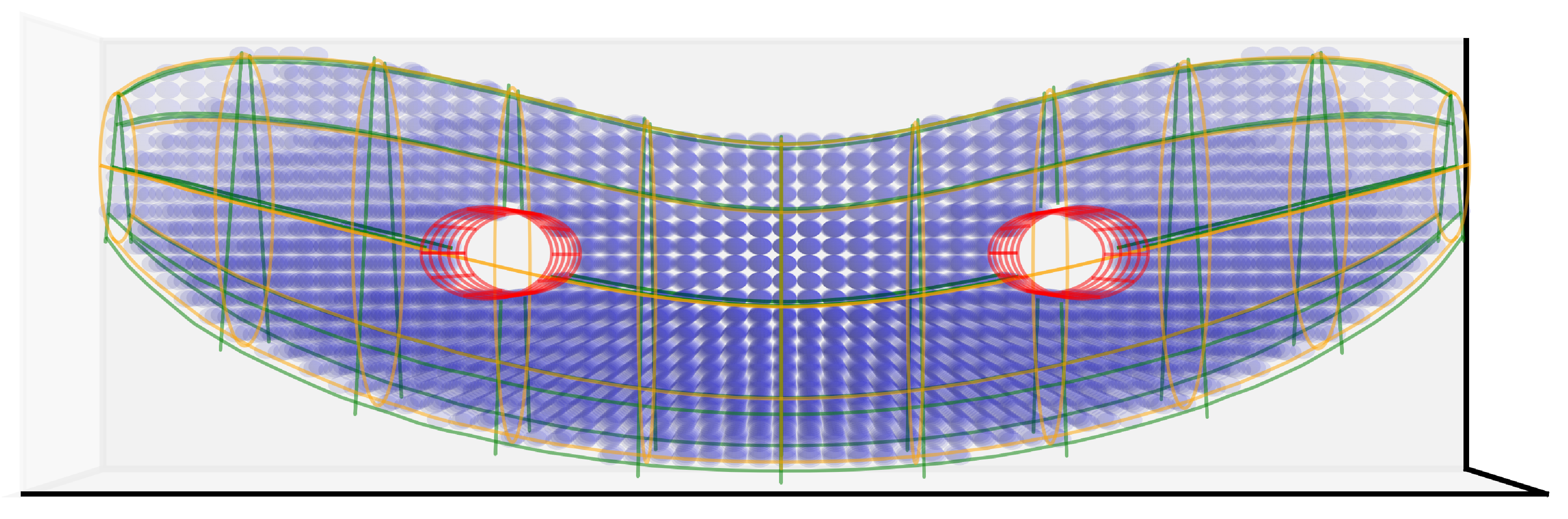}
        \includegraphics[width=0.13\textwidth]{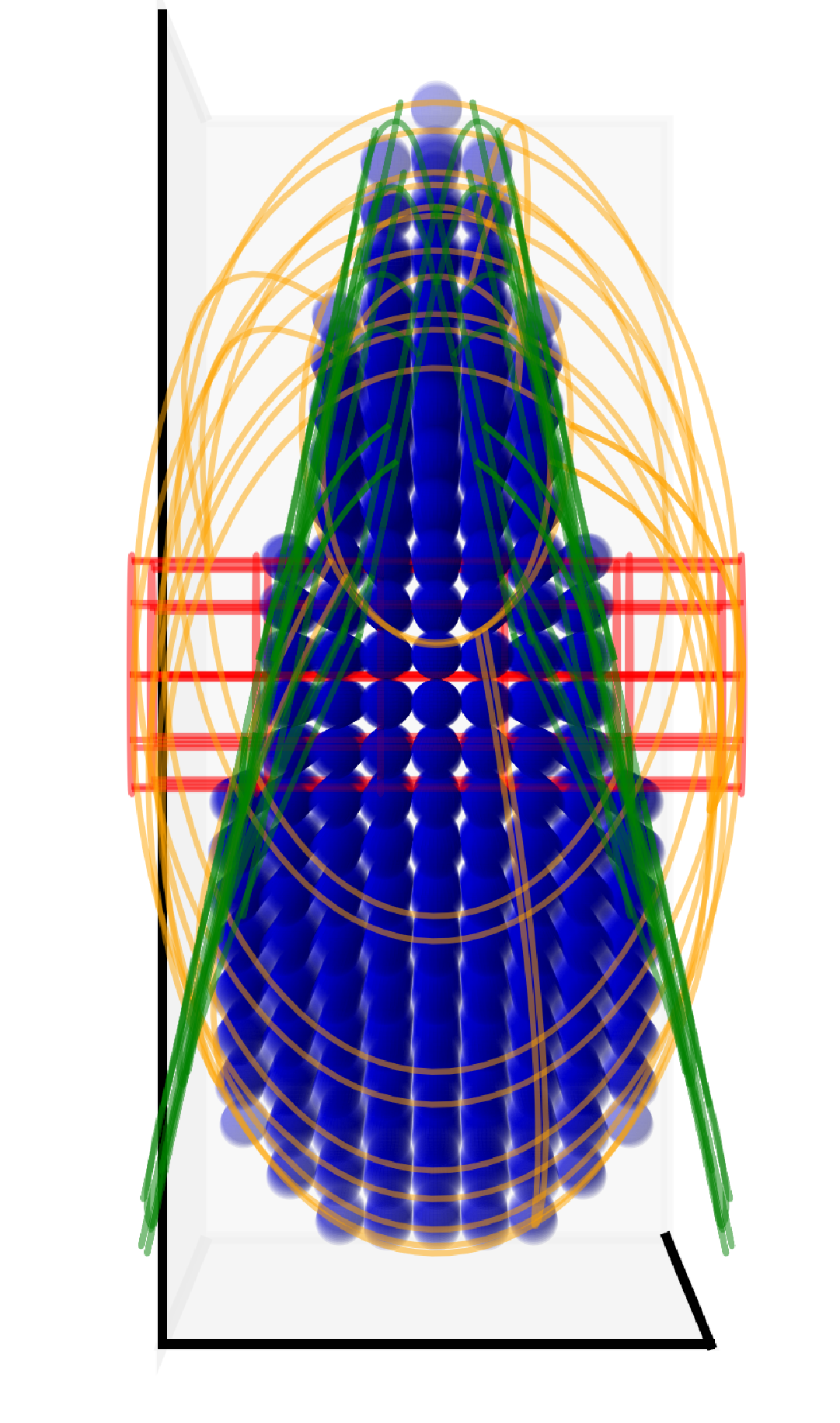}
        \caption{Crescent moon: front and side views}
        \label{fig:Moon_front}
    \end{subfigure}
    \caption{
    The bead model of the crescent moon. 
    The figure shows three auxiliary surface meshes delimiting the boundary of the modelled body (see Appendix~\ref{AppE:Moon}). 
    The crescent moon is shown in its stationary stable configuration at $\theta=0$, with $H$ along gravity.
    }
    \label{shapes_beads_moon}
\end{figure}

The translational-translational and rotational-translational mobility coefficients defined in equation~\eqref{matrices} are evaluated by the multipole expansion of the Stokes equations, corrected for lubrication to speed up the convergence. 
This method is outlined in \cite{cichocki,ekiel-jezewska2009}, and implemented in the Hydromultipole numerical codes. 
We truncate the expansion at the multipole order of $2$. 
We keep a low value of the truncation multipole order to perform computations for larger numbers of beads, and reproduce better the particle shapes. 
Typically, in this work, a particle is constructed from over 5000 spherical beads. 
For details on how the bead-model representation was created, refer to Appendix~\ref{sec:theor_param}.
Smaller numbers of beads are also used to see an approximate saturation of the mobility coefficients with the increasing number of beads inside a prescribed external surface of the particle.

The calculated values of the dimensionless mobility coefficients defined in equation~\eqref{matrices} are listed in table~\ref{tab_mob_num}. 
The numerical calculation confirms that for all the objects, $\mu_{51} \,\mu_{42}<0$. 
Therefore, based on the theoretical stability analysis performed in \cite{joshi} and \cite{ekiel2009hydrodynamic}, we conclude that for all the particle types investigated in this work, the final stationary orientation is stable. 

\begin{table}[h!]
    \centering
    \begin{tabular}{c|c|c|c|c|c|c|c|c|c}
        \hline
        Object (shape) & $L/d$ & Number of beads & $\mu_{11}$ & $\mu_{22}$ & $\mu_{33}$ & $\mu_{42}$ & $\mu_{51}$ & $|\mu_{51}/\mu_{42}|$ & $\frac{\mu_{51}+\mu_{42}}{|\mu_{51}|}$\\ \hline
        Cone          & 21 & 5088   & 0.269  & 0.269  & 0.303  & 0.057  & -0.057 & 1.00 & 0.0\\
        Crescent moon & 55 & 5513   & 0.751  & 0.560  & 0.639  & 0.381  & -0.167 & 0.44 & 1.281\\
        Arrowhead     & 29 & 5104   & 0.422  & 0.370  & 0.396  & 0.050  & -0.067 & 1.34 & -0.254\\
        Ring (1 mm)   & 67 & 5182   & 0.559  & 0.442  & 0.552  & 0.057  & -0.058 & 1.001 & -0.017 \\
        Ring (2 mm)   & 73 & 5106   & 0.608  & 0.481  & 0.592  & 0.096  & -0.102 & 1.06 & -0.059\\
        Ring (3 mm)   & 81 & 5110   & 0.671  & 0.528  & 0.643  & 0.140  & -0.145 & 1.04 & -0.034\\
        Ring (4 mm)   & 85 & 4870   & 0.717  & 0.565  & 0.679  & 0.193  & -0.195 & 1.01 & -0.010\\ 
    \end{tabular}
    \caption{The dimensionless mobility coefficients defined in \eqref{matrices}, for the bead model of the particles used in our experiments.}
   \label{tab_mob_num}
\end{table}

\section{Discussion and conclusions}\label{conclusions}
\begin{figure}
    \centering
    \includegraphics[width=0.6\linewidth]{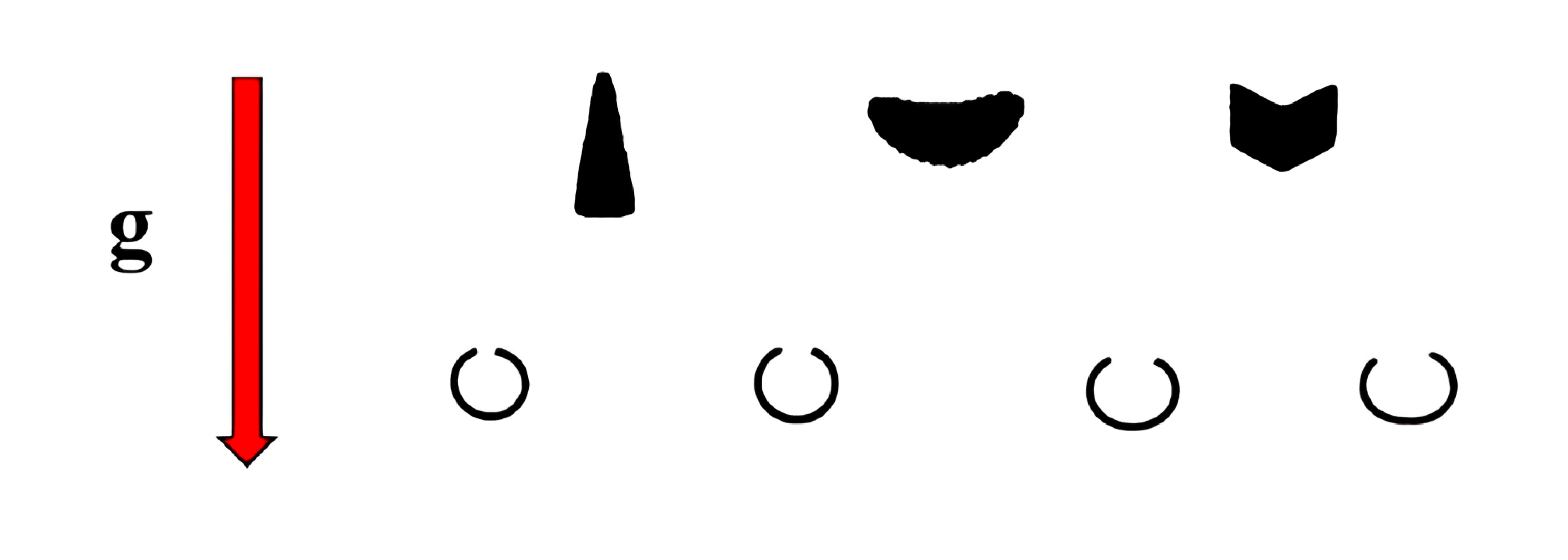}
    \caption{
    Stationary stable orientations of different objects, corresponding to $\theta=0$. 
    The shapes and their orientation are taken from the experimental images.
    } 
    \label{fig:stable_position}
\end{figure}

In this work, we studied experimentally and numerically the gravitational settling 
of rigid particles in a viscous fluid at the Reynolds number much smaller than unity. We investigated the dynamics of particles with different shapes, including cones, arrowheads, crescent moons, and open rings with four different gap sizes. All these objects had two perpendicular symmetry planes. In general, there are three families of such shapes, characterised by qualitatively different dynamics, classified in \cite{joshi}. The goal of this work was to analyse one of them: the settlers -- particles that approach a stationary orientation. Dynamics of a trumbbell, an example of the settler, was studied in \cite{ekiel2009hydrodynamic}. 
Indeed, in the experiments performed in the present work, the chosen objects reoriented and achieved stable stationary orientations, shown in figure~\ref{fig:stable_position} using experimental images of these objects. 

As described in section~\ref{thb}, the dynamics of the settlers are characterised by two rotational-translational mobility coefficients, $\mu_{51}$ and $\mu_{42}$, having opposite signs. Their values carry information about how fast the particle reorients. Therefore, in this work, we aimed to measure these coefficients in the experiments. We chose such initial orientations that, owing to symmetry, might lead to the particle rotation around $y=Y$ for the inverted initial orientation, and $x=X$ for the inclined initial orientation, as illustrated in figure~\ref{fig:sing_conv:mu51}-\ref{fig:sing_conv:mu42}. 
For the cones, arrows, and open rings, we selected some of such symmetric experimental trials for the inverted and inclined initial inclinations, and determined, respectively, $\mu_{51}$ and $\mu_{42}$, listed in table \ref{tab:mobility_velocity}. In practice, there are always certain perturbations in the experiments.  For crescent moons initially at the inverted initial orientation, the perturbations increased with time in such a way
that we could not determine the coefficient $\mu_{51}$, even though we performed a huge number of experimental trials. Instead, for crescent moons with the inverted initial orientation, we measured experimentally the coefficient $\mu_{42}$. 

The theoretical explanation follows from the stability analysis of the rotations around $y=Y$  and $x=X$, performed in \cite{Zdybel_unpublished_2025}. The result is that rotations around $y=Y$  and $x=X$ are unstable against perturbations around the $z$ axis for $(\mu_{42}+\mu_{51})\cos \theta < 0$ and $(\mu_{42}+\mu_{51})\cos \theta > 0$, respectively. The signs of $\mu_{42}+\mu_{51}$, evaluated numerically, are listed in the last column of table \ref{tab_mob_num}. For the cone, $\mu_{42}+\mu_{51}=0$, what explains the absence of increasing perturbations out of the $XZ$ plane. 

For the crescent moon, $\mu_{42}+\mu_{51}>0$. Therefore, for the initially inverted orientation, if a perturbation of $\cos \theta \approx-1$ triggers the particle motion in the $XZ$ plane (i.e, the rotation around $y=Y$), a significant instability occurs against an additional rotation around the $z$ axis. The typical time scale $1/|\mu_{51}|$ of the rotation around $y=Y$ is larger than the time scale $1/(\mu_{51}+\mu_{42})$ to rotate around $z$ and move out of the $XZ$ plane (see the last column in table \ref{tab_mob_num}), which leads to a 3-dimensional dynamics dominated by the rotation around $z$. 
An experimental example is shown in Appendix~\ref{sec:other_behaviour} in figure~\ref{fig:sedim_other_behaviour:a}, where for $\theta >\pi/2$, the crescent moon quickly rotates around the $z$ axis, moving out of the $XZ$ plane while reorienting around the $Y$ axis. For $\theta< \pi/2$, rotation around $z$ is not seen, in agreement with the stability result presented above. Some of the initial perturbations out of the inverted orientation trigger the rotation around the $x=X$ axis, which is stable against rotation around $z$ for $\theta > \pi/2$, and weakly unstable for $\theta <\pi/2$, which allows for the measurement of $\mu_{42}$. The rotation around $z$, caused by this weak instability, with the numerically evaluated $(\mu_{51}+\mu_{42})/|\mu_{42}|=0.56<1$,
is visible in figure~\ref{fig:sedim_CMA:c}. Other perturbations lead to a complex combination of different components of the angular velocity, as shown in the repository \cite{repository}.

For the arrowhead, $\mu_{42}+\mu_{51}<0$, and the rotation around $y=Y$ is stable against rotation around $z$ for $\theta > \pi/2$, but unstable for $\theta <\pi/2$. The small rotation around $z$ caused by this instability is visible in figure~\ref{fig:sedim_CMA:b}. For the shown experimental trial, the arrowhead rotated around $z$ by a small angle, much smaller than for the crescent moon. This is in agreement with the calculated and measured value  $(\mu_{51}+\mu_{42})/|\mu_{51}|=0.25$, which for the arrowhead is 5 times smaller than for the crescent moon rotating around $y=Y$ and 2 times smaller than for the crescent moon rotating around $x=X$. 

The interesting observation is that the rotation around $z$, i.e., a chiral motion, takes place for the non-chiral shapes: crescent moon and arrowhead. This effect is caused by non-vanishing rotational-translational mobility coefficients, resulting in the coupling of rotational and translational motion. More complex chiral motion has been observed for the non-chiral shapes with two orthogonal symmetry planes, called flutterers, see \cite{miara2024dynamics,vaquero2024u,joshi,vaquero2025fluttering}.

For the open rings, $|\mu_{42}+\mu_{51}|$ is small, significantly smaller than the experimental uncertainty. In addition, the rings are opened manually, which introduces some deviations from symmetry. These small asymmetries may be the main reason for the 3-dimensional dynamics observed for some trials, as shown in Appendix~\ref{sec:other_behaviour} in figure~\ref{fig:sedim_other_behaviour:b}-\ref{fig:sedim_other_behaviour:c}. 

In general, there is a good agreement between the mobility coefficients measured in the experiments (listed in table~\ref{tab:mobility_velocity}) and the mobility coefficients evaluated numerically for the theoretical bead model of the same shape (shown in table~\ref{tab_mob_num}). Therefore, the theoretical model presented in this work may serve as a useful tool to determine values of $\mu_{51}$ and $\mu_{42}$ for objects of the required shapes, without performing experiments. These values allow predicting the type of dynamics, and in particular, to check if an object is a settler, and how fast it will rotate to the stable orientation. This information is important for practical applications. 

Our results have been obtained for millimetre-sized particles settling in a highly viscous fluid at the Reynolds number much smaller than unity. Owing to the similarity principle (see, e.g., \cite{Homsy}), the obtained dynamics can be easily rescaled to the motion of geometrically similar microobjects in a water-based solution, at the same value of the Reynolds number. 

Finally, our results indicate that a dilute suspension of sedimenting micro-settlers after a certain time will form an ordered, anisotropic medium. Such a system is of theoretical and practical interest.

\section*{Acknowledgements} \label{ackn}
\textit{
    This work was supported in part by the National Science Centre under grant UMO-2021/41/B/ST8/04474. 
}

\section*{Declaration of interests}
\textit{
    The authors report no conflict of interest.
}

\section*{Data availability statement}
\textit{
    The data that support the findings of this study are openly available in RepOD -- Repository for Open Data at \url{https://doi.org/10.18150/MIL6UG}. 
}

\appendix

\section*{Appendices}

\section{3-dimensional reorientation of a sedimenting particle} 
\label{sec:other_behaviour}
\begin{figure}
    \centering
    \begin{subfigure}{\textwidth}
        \centering
        \includegraphics[clip, trim=2.1cm 2.1cm 2.1cm 2.1cm, width=0.9\linewidth]{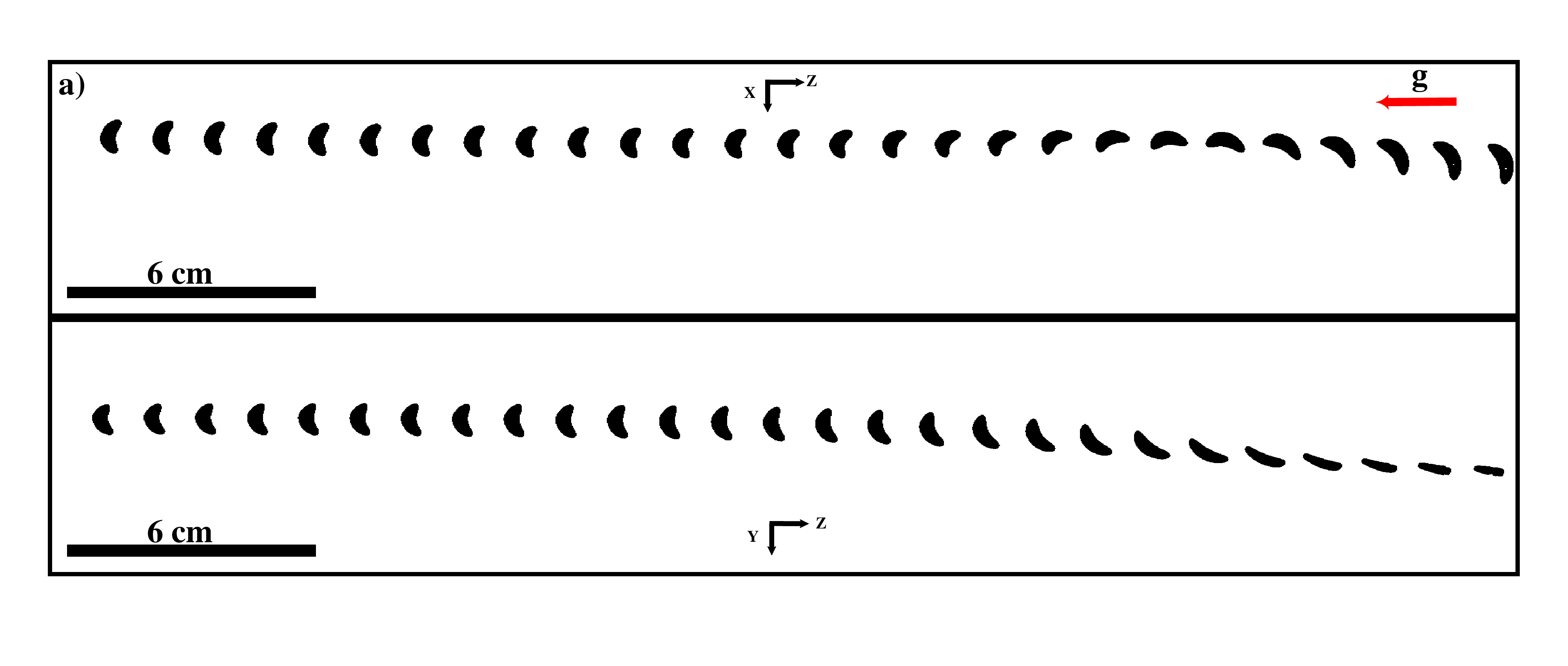} 
        \phantomsubcaption
        \label{fig:sedim_other_behaviour:a}
    \end{subfigure}
    \begin{subfigure}{\textwidth}
        \centering
        \includegraphics[clip, trim=2.1cm 2.1cm 2.1cm 2.1cm, width=0.9\linewidth]{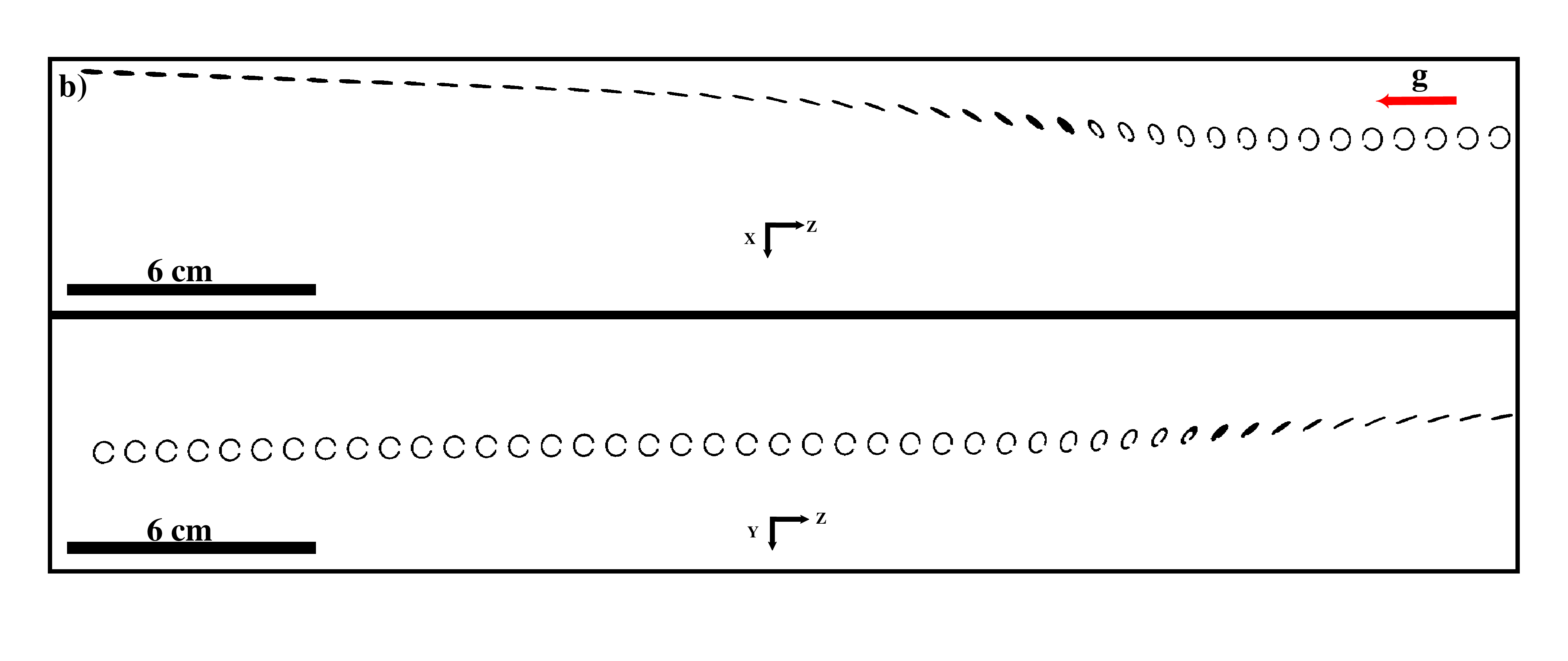} 
        \phantomsubcaption
        \label{fig:sedim_other_behaviour:b}
    \end{subfigure}
    \begin{subfigure}{\textwidth}
        \centering
        \includegraphics[clip, trim=2.1cm 2.1cm 2.1cm 2.4cm, width=0.9\linewidth]{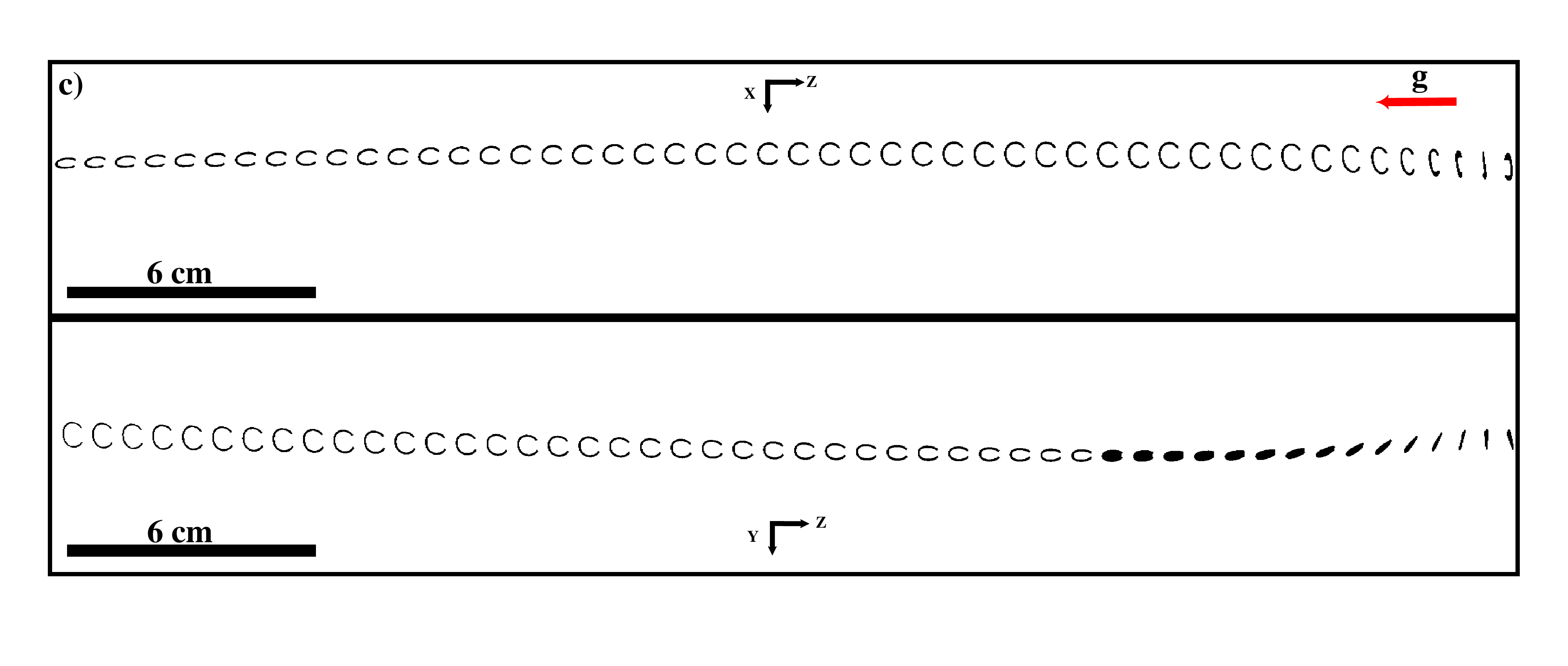} 
        \phantomsubcaption
        \label{fig:sedim_other_behaviour:c}
    \end{subfigure}
    \caption{Snapshots from 3 experimental trials, demonstrating 3-dimensional reorientation of a single sedimenting object: 
        a) crescent moon in the inverted initial orientation,
        b) ring with the opening of 1~mm in the inverted initial orientation, and
        c) ring with the opening of 4~mm in the inclined initial orientation.
        The images were recorded simultaneously by two cameras (top and bottom panels). 
        Gravity points left, and the particles move from right to left.
    }
    \label{fig:sedim_other_behaviour}
\end{figure}

Figure~\ref{fig:sedim_other_behaviour} presents examples of 3-dimensional reorientation of a sedimenting particle initially at the ``inverted'' or ``inclined'' orientations. Such a complex dynamics have not been analysed quantitatively in this work. It is evident that for each trial shown in figure~\ref{fig:sedim_other_behaviour}, the time-dependent, non-zero components $X$, $Y$, and $Z$ of the angular velocity are present in some ranges of time.

\section{Number of experiments}\label{number}

\begin{table}
    \centering
    \begin{tabular}{c|c|c|c|c|c|c}
      Object & Stationary & \multicolumn{2}{c|}{Inverted} & \multicolumn{2}{c|}{Inclined} & Total \\  
      &  
      & Selected 
      & Total 
      & Selected 
      & Total 
      &  \\
     C & 4 & 4 &  7 & 0 &  0 & 11 \\
     M & 4 & 7 & 27 & 0 &  0 & 31 \\
     A & 3 & 7 &  9 & 6 & 10 & 22 \\
    R1 & 4 & 3 &  5 & 5 &  7 & 16 \\
    R2 & 4 & 3 &  4 & 3 &  8 & 16 \\
    R3 & 4 & 4 &  9 & 4 & 13 & 26 \\
    R4 & 4 & 3 &  5 & 3 &  7 & 16 \\  
    \end{tabular}
    \caption{
    The number of experimental trials performed for the indicated initial orientations of the objects, along with the number of trials selected for quantitative analysis of the reorientation.
    Names of the objects: C - cone, A - arrowhead, M - crescent moon, R$i$ - ring with the opening width equal to $i = 1, 2, 3, 4$ mm. 
    }
    \label{ne}
\label{c}
\end{table}

In table~\ref{ne} we list the numbers of the experimental trials performed with each type of particle, for all the initial orientations: stationary, inverted, and inclined. 
In 111 experimental trials, we observed the particle reorienting to a stable, stationary orientation (66 trials for the inverted initial orientation and 45 trials for the inclined initial orientation). In addition, 27 trials showed particles of various types remaining at a stationary, stable orientation throughout the entire experiment. 
Therefore, in this work, we performed 138 experimental trials.

\section{Measurement of the Orientation Angle $\theta$ from a 2D image}\label{extracting}

\subsection{Orientation Angle Estimation Using Second Central Moments of the Binary Image}\label{app:extracting:1}

We employ a statistical method based on second central moments to determine an object's orientation angle in a binary image, as in \cite{BurgerBurge2009}. The normalized central moments of order $(p+q)$ are defined by
\bee
\mathcal{M}_{pq} = \frac{1}{N_{\mathcal{B}}^{(p+q+2)/2}} \sum_{i \in \mathcal{B}} (x_i - \bar{x})^p (z_i - \bar{z})^q,
\eee
where $(x_i, z_i)$ denotes the coordinates of the pixels comprising the analyzed 2D object $\mathcal{B}$, and
$$(\bar{x}, \bar{z})=\frac{1}{N_\mathcal{B}}\sum_{i\in \mathcal{B}} (x_i, z_i)$$ 
is the centroid of $\mathcal{B}$, and $N_{\mathcal{B}}$ is the total number of pixels in $\mathcal{B}$.

\begin{figure}
    \centering
    \includegraphics[width=0.3\linewidth]{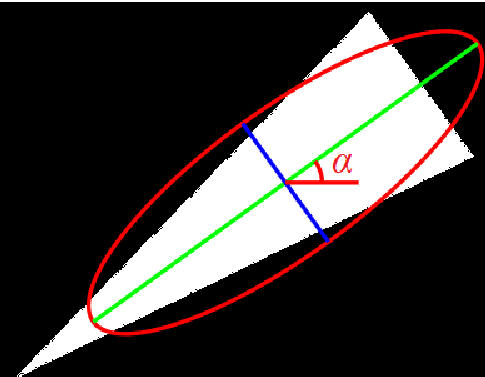}
    \caption{
        A schematic binary image with the computed equivalent ellipse and the indicated inclination angle $\alpha$. 
        The ellipse is centred at the object's centroid, and its principal axes are also marked.
    }
    \label{fig:bw_image_an}
\end{figure}

The covariance matrix $\boldsymbol{\Sigma}$ is constructed using the second moments of $\mathcal{B}$:
\bee
\boldsymbol{\Sigma}=\left(\bt{ll} $\mathcal{M}_{20}$ & $\mathcal{M}_{11}$ \\
$\mathcal{M}_{11}$ & $\mathcal{M}_{02}$ \et
\right).
\label{eq:Sigma_Cov}
\eee
The above matrix is diagonalised using a similarity transformation $\tilde{\boldsymbol{\Sigma}}= \boldsymbol{R}^{-1}\boldsymbol{\Sigma} \boldsymbol{R}$, where $\boldsymbol{R}$ is a rotation matrix, which rotates counterclockwise the coordinate axes by an angle $\alpha$ given by
\bee
\alpha = \frac{1}{2}\arctan\left(\frac{2 \mathcal{M}_{11}}{\mathcal{M}_{20}-\mathcal{M}_{02}}\right),
\eee
where the orientation angle of the object $\mathcal{B}$ used in our measurements is equal to 
\bee
\theta= \frac{\pi}{2} + \alpha,
\eee
where both angles are measured in radians. The angle $\alpha$ for a given body $\mathcal{B}$ is depicted in figure~\ref{fig:bw_image_an}.

\subsection{Equivalent Ellipse of the 2D Object}\label{app:extracting:2}

The object  $\mathcal{B}$ under study can be represented by an equivalent ellipse with a major semi-axis of length $a$ and a minor semi-axis of length $b$, which exhibits the same covariance matrix as the object  $\mathcal{B}$ (see figure~\ref{fig:bw_image_an}). Diagonalisation of the covariance matrix ensures that, after a rotation by an angle $\alpha$, the major axis of the ellipse is horizontal, and the covariance matrix is given by $\tilde{\boldsymbol{\Sigma}}=\mathrm{diag}(a^2,b^2)$, where 
$$a^2=\frac{1}{2}\left(\mathcal{M}_{20}+\mathcal{M}_{02}+\sqrt{(\mathcal{M}_{20}-\mathcal{M}_{02})^2+4\mathcal{M}_{11}^2}\right)$$ and 
$$b^2=\frac{1}{2}\left(\mathcal{M}_{20}+\mathcal{M}_{02}-\sqrt{(\mathcal{M}_{20}-\mathcal{M}_{02})^2+4\mathcal{M}_{11}^2}\right).$$

\subsection{Uncertainty of the Orientation Angle}\label{app:extracting:3}

We estimate the measurement uncertainty of the object's orientation angle $\theta=\pi/2 + \alpha$ using the propagation of measurement error
\bee
\sigma_{\theta}=\sqrt{\left(\frac{\partial \alpha}{\partial \mathcal{M}_{20}}\sigma_{\mathcal{M}_{20}}\right)^2+\left(\frac{\partial \alpha}{\partial \mathcal{M}_{02}}\sigma_{\mathcal{M}_{02}}\right)^2+\left(\frac{\partial \alpha}{\partial \mathcal{M}_{11}}\sigma_{\mathcal{M}_{11}}\right)^2},
\label{eq:prop_err}
\eee
where $\frac{\partial \alpha}{\partial \mathcal{M}_{02}}=-\frac{\partial \alpha}{\partial \mathcal{M}_{20}}=2\mathcal{M}_{11} D^{-1}$, $\frac{\partial \alpha}{\partial \mathcal{M}_{02}}=(\mathcal{M}_{20}-\mathcal{M}_{02}) D^{-1}$ and $D=(\mathcal{M}_{20}-\mathcal{M}_{02})^2+4\mathcal{M}_{11}^2$. We seek an estimate of the measurement uncertainty of the angle $\theta$, assuming that pixel coordinates are random variables $(X, Z)$, which follow a bivariate normal elliptical distribution $\mathcal{N}_2((0,0), \boldsymbol{\Sigma}),$ where the covariance matrix is the same as in equation~\eqref{eq:Sigma_Cov} for this distribution, but we use a convenient parametrization equal to
\bee
\boldsymbol{\Sigma}=\boldsymbol{R}\tilde{\boldsymbol{\Sigma}}\boldsymbol{R}^{-1}=\left(\bt{ll} $a^2 \cos^2\alpha+b^2 \sin^2\alpha$ & $(a^2-b^2)\sin\alpha\cos\alpha$ \\
$(a^2-b^2)\sin\alpha\cos\alpha$ & $a^2 \sin^2\alpha+b^2 \cos^2\alpha$ \et
\right),
\eee
where $\alpha=\theta-\pi/2$ is related to the measured value of the orientation angle $\theta$, $a$ and $b$ are semi-axes of the equivalent ellipse of the object $\mathcal{B}$. We apply Isserlis' theorem (see \cite{Isserlis1916, Isserlis1918}) to expressions for the expected values of standard deviations for the second central moments of $\mathcal{B}$, which leads to
\begin{align}
\sigma^2_{\mathcal{M}_{20}}&=\frac{2}{N_{\mathcal{B}}}\left(a^2 \cos^2\alpha+b^2 \sin^2\alpha\right)^2,\\
\sigma^2_{\mathcal{M}_{02}}&=\frac{2}{N_{\mathcal{B}}}\left(a^2 \sin^2\alpha+b^2 \cos^2\alpha\right)^2,\\
\sigma^2_{\mathcal{M}_{11}} &=\frac{1}{4 N_\mathcal{B}}\left[\left(a^2+b^2\right)^2-\left(a^2-b^2\right)^2 \cos4\alpha\right] \nonumber.
\end{align}
We use the above expressions and equation~\eqref{eq:prop_err}, to obtain
\bee
\sigma_\theta=\frac{|\cos 2\alpha|}{|a^2-b^2|}\sqrt{\sigma^2_{\mathcal{M}_{11}}+\frac{1}{4}\tan^2 2\alpha\left[\sigma^2_{\mathcal{M}_{20}}+\sigma^2_{\mathcal{M}_{02}}\right]}.
\eee
As our final error estimation, we use the value for $\alpha = 0$ equal to
\bee
\sigma_{\theta, est}=\frac{1}{\sqrt{N_\mathcal{B}}}\frac{ab}{|a^2-b^2|}.
\eee
This expression slightly underestimates the error, but the differences are minor in the case of the rigid bodies studied in this experiment. It is worth noting that the object's orientation becomes undefined for objects with nearly equal semi-axes (i.e., $a\approx b$). We can straightforwardly write down the expression for $\sigma_{\tan\frac{\theta}{2}}$, which is given by
\bee
\sigma_{\tan\frac{\theta}{2}}=\left|\frac{\mathrm{d} \tan\frac{\theta}{2}}{\mathrm{d}\theta}\right|\sigma_{\theta, est}=\frac{1}{2\sqrt{N_\mathcal{B}}}\frac{ab}{|a^2-b^2|}\left(1+\tan^2\frac{\theta}{2}\right).
\label{eq:ang_err}
\eee
The above formula was used to compute error bars in figure~\ref{tan_theta_vs_t}.

Image analysis was performed using MATLAB's built-in function {\tt regionprops}, which provides direct access to the properties of the binary image. Notably, this function allows one to obtain the angle $\alpha$ through the {\tt "Orientation"} property, as well as the lengths of the axes of the equivalent ellipse using the {\tt "MajorAxisLength"} and {\tt "MinorAxisLength"} properties.

\subsection{Irrelevance of the Motion Blur}
Due to the finite exposure times of the cameras used in the experiments, motion blur occurs as a result of the body's vertical sedimentation. We assume that the body's horizontal velocity is negligible compared to its sedimentation velocity, thereby allowing the motion blur effect to be considered solely in the $z$-direction. The standard deviation $\sigma_{blur}$ associated with this effect is obtained by modelling the blur as a uniform distribution along the $z$-axis with a width of $UTk$, which leads to
\bee
\sigma^2_{blur} = \frac{(UTk)^2}{12}.
\eee
where 
$U$ is the sedimentation velocity, $T$ is the exposure time, and $k$ is a scaling factor that converts distances measured in millimetres to pixel units, depending on the camera calibration. 

Motion blur alters the value of the second-order central moment $\mathcal{M}_{20}$, typically leading to its overestimation due to the elongation of the object's intensity profile along the direction of blur. The relationship between the second-order moments in the presence of motion blur and its absence is given by:
\bee
\mathcal{M}^{blur}_{20}&=&\mathcal{M}_{20}, \nonumber\\
\mathcal{M}^{blur}_{02}&=&\mathcal{M}_{02}+ \sigma^2_{blur}, \\
\mathcal{M}^{blur}_{11}&=&\mathcal{M}_{11}, \nonumber
\eee
Consequently, the orientation angle of the body is affected. We can relate the true orientation angle $\alpha$ to the one extracted from a blurred image, $\alpha_{blur}$, through the following equation:
\bee
\tan2 \alpha = \frac{\tan2 \alpha_{blur}}{1+\frac{\sigma^2_{blur}}{2 \mathcal{M}_{11}} \tan2 \alpha_{blur}}\approx\frac{\sin2 \alpha_{blur}}{\cos 2 \alpha_{blur}+\frac{\sigma^2_{blur}}{a^2_{blur}-b^2_{blur}}},
\eee
where $a_{blur}$ and $b_{blur}$ are the lengths of the semiaxes of the equivalent ellipse obtained from the blurred image. In the absence of motion blur, the estimated orientation angle coincides with the true orientation angle, i.e., $\alpha=\alpha_{blur}$.
For the objects studied $\sigma_{blur}^2/\left(a_{blur}^2-b_{blur}^2\right)\ll 1$, because $\sigma^2_{blur}$ is of the order of a few squared pixels, whereas $a_{blur}^2-b_{blur}^2$  is of the order of thousand squared pixels for all the objects. The relative difference between the values of $\tan2\alpha$ is given by
\bee
\frac{|\tan 2\alpha_{blur}-\tan 2\alpha|}{|\tan 2\alpha_{blur}|}\approx \frac{\sigma_{blur}^2}{a_{blur}^2-b_{blur}^2}\frac{1}{\cos 2\alpha_{blur}}
\eee
and is smaller than $1\%$ for almost all angles except those equal to $45^\circ \pm 2^\circ$. Therefore, the effect of motion blur can be safely neglected in the present study.

\section{Parameters of the fit equation~\eqref{eq:exp_exp} to the experimental data in figure~\ref{tan_theta_vs_t} }
\label{sec:experimental_parameters}

Table \ref{tab:fitting_param} gives the fitting parameters of theoretical function 
\eqref{eq:exp_exp} to the data shown in figure~\ref{tan_theta_vs_t} for seven experimental trials with particles of various shapes. The respective values of the coefficient of determination $R^2$ indicate the good agreement between the experiment and the theoretical prediction. In our study, the coefficient of determination $R^2$ was calculated as
\bee
R^2=1-\frac{\sum_{i=1}^n w_i \left(y_i - \hat{y}_i\right)^2}{\sum_{i=1}^n w_i \left(y_i - \bar{y}_w\right)^2},
\eee
where $y_i=\tan(\theta_i/2)$ are the measured data at time instance $\tilde{t}_i$, $\hat{y}_i=A \exp(-\tilde{t}_i/\tilde{\tau})$ are the model predictions with fitting parameters $A$ and $\tilde{\tau}$, $w_i = \sigma^{-2}_{\tan(\theta_i/2)}$ are the statistical weights associated with the uncertainty in equation~\eqref{eq:ang_err} and 
\bee
\bar{y}_w=\frac{\sum_{i=1}^n w_i y_i}{\sum_{i=1}^n w_i},
\eee
is the weighted average of the data points. Here, $n$ denotes the total number of measurements. Parameter estimation was carried out in MATLAB using the weighted nonlinear least squares method.

\begin{table}
    \centering
    \begin{tabular}{c|c|c|c}
        Object (shape)   
        & A 
        & 
        $1/\tilde{\tau}$ (s$^{-1}$) 
        & $R^2$ \\ 
        Cone             &     2.743 $\pm$ 0.002  &         0.1244 $\pm$ 0.0001  & 1.000 \\
        Crescent moon    &     2.462 $\pm$ 0.040  &         0.1643 $\pm$ 0.0022  & 0.999 \\
        Arrowhead        &     5.776 $\pm$ 0.314  &         0.2136 $\pm$ 0.0083  & 0.989 \\
        Open ring (1 mm) &     2.865 $\pm$ 0.043  &         0.0407 $\pm$ 0.0005  & 0.992 \\
        Open ring (2 mm) &     2.048 $\pm$ 0.010  &         0.0599 $\pm$ 0.0003  & 0.999 \\
        Open ring (3 mm) &     1.539 $\pm$ 0.012  &         0.0639 $\pm$ 0.0005  & 0.998 \\
        Open ring (4 mm) &     2.432 $\pm$ 0.086  &         0.0604 $\pm$ 0.0021  & 0.971 \\
    \end{tabular}
    \caption{
    The fitting parameters $A$ and $1/\tilde{\tau}$ of the exponential decay fit \eqref{eq:exp_exp} to the data from seven experimental trials plotted in figure~\ref{tan_theta_vs_t} for various shapes, together with the coefficient of determination $R^2$ that shows the goodness of the fit.
    }
    \label{tab:fitting_param}
\end{table}

\section{Bead model of the particles used in the experiments}
\label{sec:theor_param}

\subsection{Cone}
The cone-shaped particle used in the experiments was in the numerical calculation approximated by a cone with the base diameter $L$ and the height $H$ equal to the corresponding experimental values specified in section~\ref{sec:exp_sys:rigid_particles}.
The bead-model representation of the cone-shaped particle was constructed using the parametric equation $R(z) = (H-z)L/(2H)$  
that describes the cone boundary  by relating its radial and azimuthal coordinates $R$ and $z.$
The spheres were positioned in a regular 3D cubic lattice with a lattice constant equal to the sphere diameter $d$, with the following constraint: the radial and azimuthal components $r$ and $z$ of the sphere centre position should satisfy the constraint $r \leq R(z)$ for $0 \leq z \leq  H$, ensuring all sphere centres remained strictly within the conical volume, including its surface. 

The sphere diameter was systematically decreased until an approximate saturation of the mobility coefficients was reached.
The resulting bead-model representation of the cone-shaped object consists of 5088 spheres and is shown in figure~\ref{fig:Cone}.
\begin{figure}[h!]
    \centering
    \includegraphics[width=0.25\linewidth]{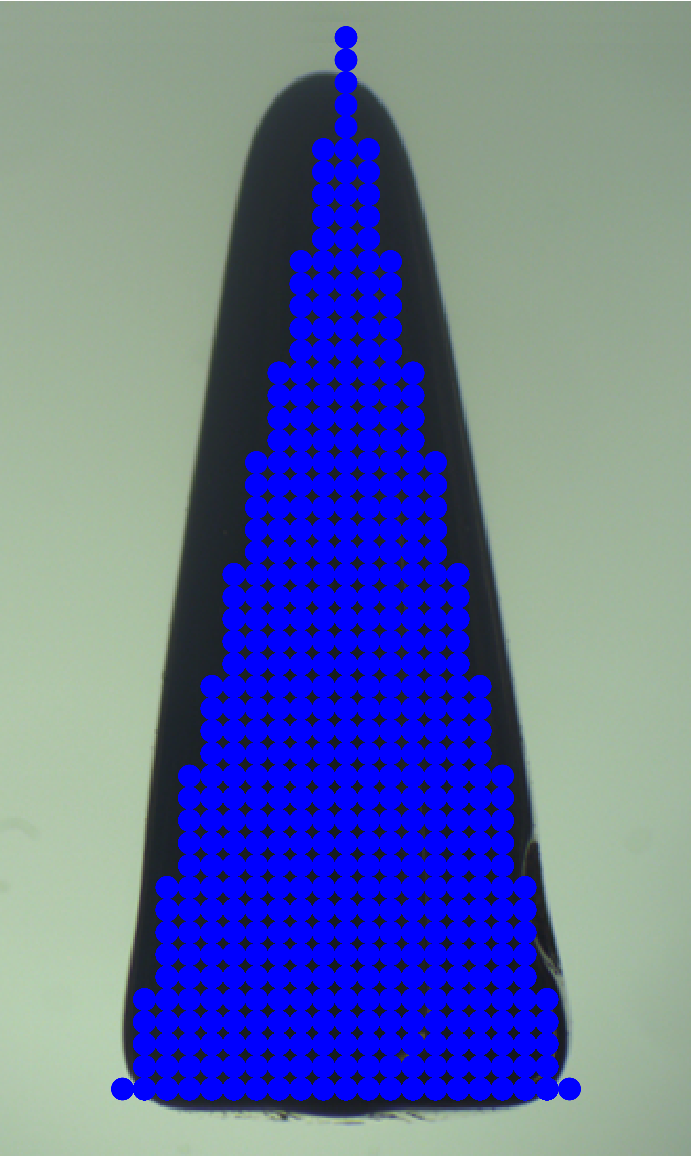}
    \caption{
    Experimental image of a cone (grayscale) overlaid with its bead-model parametrization (blue spheres), which was used for computational analysis. 
    }
    \label{fig:cone_exp_vs_sim}
\end{figure}

Note that this parametrisation leaves some regions of the real cone-shaped object not being fully represented (see figure~\ref{fig:cone_exp_vs_sim}), which results in some differences in values of the rotational-translational mobility coefficients, but not in their signs. Therefore, the rotational-translational mobility coefficients for the experimental and numerical shapes still satisfy the same essential constraint  $\mu_{51}\,\mu_{42}<0$, which means that they rotate towards a stable, stationary orientation while settling under gravity.

\subsection{Arrowhead}

In order to construct a bead-model representation of the arrowhead-shaped object used in the experiments, a rectangular prism (cuboid) was first created that had a length of $L$, a height of $H$ and a width of $W$, which corresponded to the length, total height, and width of the arrowhead-shaped object (experimental values were specified in section~\ref{sec:exp_sys:rigid_particles}).
The prism was positioned to satisfy the following constraints: $-L/2 \leq x \leq L/2$, $0 \leq y \leq W$, and $0 \leq z \leq H$.
Then, a regular 3D cubic lattice with a lattice constant equal to the sphere diameter $d$ was created such that one of the lattice nodes has coordinates $(0,0,0)$.
Finally, the model was constructed to include only the spheres whose centres at $(x,y,z)$ lie within the object's volume, which is defined by the following constraints:
\begin{equation}
    0 \leq |x| \leq L/2, \qquad 
    0 \leq y \leq W, \qquad 
    k \cdot |x| \leq z \leq S + k \cdot |x|
\end{equation}
where the slope $k$ is defined using the middle height $S$ of the arrowhead-shaped object: 
$k=\frac{H-S}{L/2}$.
The sphere diameter was systematically decreased until approximate saturation of values of all the mobility coefficients was reached.

The resulting bead-model representation of the arrowhead-shaped object consists of 5104 spheres and is shown in figure~\ref{fig:Arrow}.

\subsection{Open Ring}

We consider a rigid toroidal body in the shape of a circular pipe of radius \( R > 0 \) centred along a planar elliptical curve. The directrix of the pipe is the planar ellipse given by
\[
\mathcal{C} = \left\{ \mathbf{r}(\vartheta) = \left( a\cos\vartheta,\; b\sin\vartheta,\; 0 \right) \;\middle|\; \vartheta \in [0, 2\pi) \right\},
\]
with semi-major axis $a$ and semi-minor axis $b$. The values of both $a$ and $b$ for the ring under consideration were determined from experimental measurements.  The ellipse lies entirely in the plane \( z = 0 \), and the parameter \(\vartheta\) increases counterclockwise.

At each point on the ellipse, we define a local orthonormal frame \( \{ \mathbf{T}(\vartheta), \mathbf{N}(\vartheta), \mathbf{B}(\vartheta) \} \), where \( \mathbf{T}(\vartheta) \) is the unit tangent vector, \( \mathbf{B}(\vartheta) = \mathbf{e}_z \) is the unit vector normal to the ellipse plane, and \( \mathbf{N}(\vartheta) = \mathbf{B}(\vartheta) \times \mathbf{T}(\vartheta) \) is the in-plane normal. The pipe   surface is then parameterised as:
\[
\mathcal{S} = \left\{ \mathbf{X}(\vartheta, \varphi) = \mathbf{r}(\vartheta) + R \left[ \cos\varphi\, \mathbf{N}(\vartheta) + \sin\varphi\, \mathbf{B}(\vartheta) \right] \;\middle|\; \vartheta \in [0,2\pi),\; \varphi \in [0,2\pi) \right\}.
\]

We discretise the volume enclosed by the pipe surface using a bead model. The positions of bead centres (normalised by the bead diameter) are placed on a uniform cubic grid with spacing 1, and only the beads with centres  \((x, y, z) \in \mathbb{R}^3\) that lie inside the pipe are retained. A bead centre is considered inside the pipe if its distance to the elliptical curve \(\mathbf{r}(\vartheta)\) is less than or equal to the pipe radius \(R\), i.e.
\[
\min_{\vartheta \in [0, 2\pi)} \left\| (x, y, z) - \mathbf{r}(\vartheta) \right\| \leq R.
\]
This condition is evaluated numerically by discretising the parameter \(\vartheta\) and finding the closest point on the ellipse for each grid point.

To construct the open rings similar to those used in the experiments, we remove a subset of beads located near a prescribed angular sector on the ellipse. Specifically, we define a central angular position \(\vartheta_0 = 90^\circ\), and remove all beads for which the nearest to their centre point on the ellipse corresponds to an angle \(\vartheta^* \in [\vartheta_0 - \Delta\vartheta,\; \vartheta_0 + \Delta\vartheta]\).

To ensure that the gap in the ring corresponds to a physically meaningful arc length \( \ell_\mathrm{gap} \), we determine the corresponding angular width \(\Delta\vartheta\) such that:
\[
\int_{\vartheta_0 - \Delta\vartheta}^{\vartheta_0 + \Delta\vartheta} 
\left\| \frac{\mathrm{d}\mathbf{r}}{\mathrm{d}\vartheta}(\vartheta) \right\| \, \mathrm{d}\vartheta 
= \ell_\mathrm{gap}.
\]
This is a nonlinear equation for \(\Delta\vartheta\), since the arc-length density
\[
\left\| \frac{\mathrm{d}\mathbf{r}}{\mathrm{d}\vartheta}(\vartheta) \right\| = 
\sqrt{a^2 \sin^2\vartheta + b^2 \cos^2\vartheta}
\]
depends on \(\vartheta\). We solve this equation numerically using the bisection method. Given a target arc length \(\ell_\mathrm{gap}\), the corresponding angular half-width \(\Delta\vartheta\) is determined such that the elliptical arc from \(\vartheta_0 - \Delta\vartheta\) to \(\vartheta_0 + \Delta\vartheta\) matches the required length. This procedure guarantees rotational symmetry of the gap around \(\vartheta_0\) and ensures precise geometric control over the opening.

 The ratio of the principal semi-axes $a$ and $b$ was correlated with the experimentally measured values, i.e. $(a+R)/(b+R) \approx L/H$. Similarly, the ratio of the major semi-axis to the pipe radius was correlated with $(a+R)/R\approx L/W$. 
 During the computations, the pipe surface was rescaled, which enabled control over the number of beads $N$ contained within that surface. Increasing the value of $R$ (measured in the sphere diameters $d$) rescales the pipe surface, thereby increasing the number of spheres incorporated in the model, and therefore also its accuracy. For rings with different  gap widths, the following parameters were employed to match shapes of the open rings used in the experiments and specified in section~\ref{sec:exp_sys:rigid_particles}:
\begin{itemize}
  \item[(a)] \(a = 30.90\), \(b = 27.90\), \(R = 3.6\)  and \(\Delta \vartheta = 10^\circ 47'\) for a ring with a gap width of 1 mm, which yielded \(N = 5182\);
  \item[(b)] \(a = 33.72\), \(b = 27.60\), \(R = 3.6\), and \(\Delta \vartheta = 20^\circ 32'\) for a ring with a gap width of 2 mm, which yielded \(N = 5106\);
  \item[(c)] \(a = 37.08\), \(b = 27.48\), \(R = 3.6\), and \(\Delta \vartheta = 28^\circ 30'\) for a ring with a gap width of 3 mm, which yielded \(N = 5110\);
  \item[(d)] \(a = 39.06\), \(b = 27.36\), \(R = 3.6\), and \(\Delta \vartheta = 36^\circ 35'\) for a ring with a gap width of 4 mm, which yielded \(N = 4870\).
\end{itemize}
All parameter values are expressed in units of the sphere diameter $d = 1$.

\subsection{Crescent Moon}
\label{AppE:Moon}
We constructed the crescent-shaped body used in the numerical simulations as an assembly of touching spheres of diameter~$d=1$, arranged on a regular cubic lattice. The lattice points corresponding to the centres of the spheres were selected based on a parametric description of a planar reference curve in the $xz$-plane and a rule for generating smoothly varying elliptical cross-sections in the perpendicular $yz$-plane.

\begin{figure}
    \centering
    \includegraphics[width=0.8\linewidth]{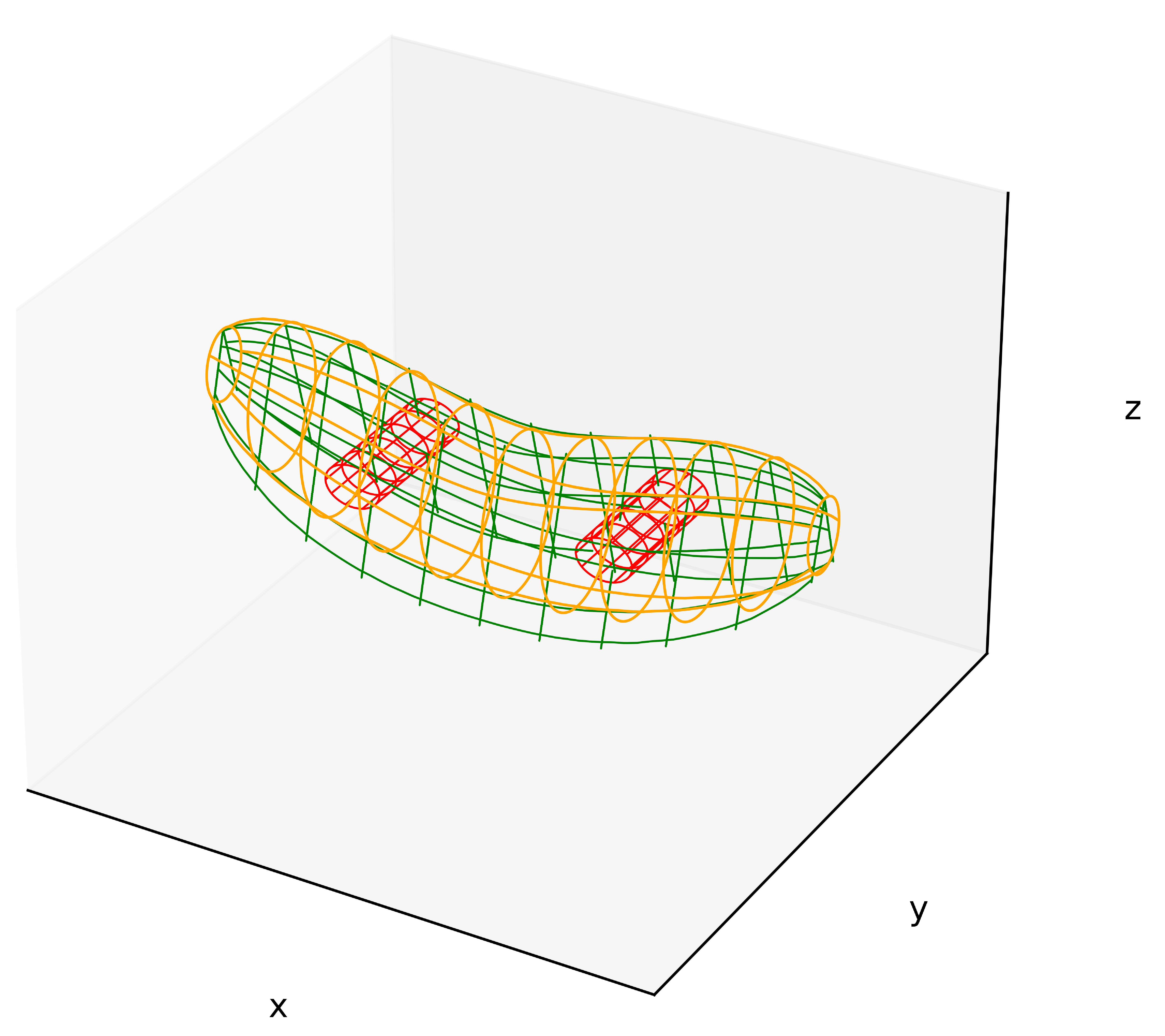}
    \caption{The surfaces meshes used in the construction of the bead model for the crescent moon.}
    \label{fig:surf_mesh}
\end{figure}
 
The construction begins with a reference curve $\Gamma$ defined in the $xz$-plane at $y = 0$, given by the parametric equations
\begin{equation}
x(p) = \eta_x \left(5\cos p - \sin 2p\right), \qquad
z(p) = \eta_z \left(3 \sin p + 2 \cos 2p\right),
\end{equation}
for $p \in [0, 2\pi)$, where $\eta_x = 0.84\,s$ and $\eta_z = 0.51\,s$, with $s > 1$ denoting a dimensionless scaling factor. The values of $\eta_x$ and $\eta_z$ were chosen to match the shape of a physical object used as a reference, i.e., the ratio $L/H$ is approximately equal for experimental measurements and numerics. For each integer multiple of $d$ along the $x$-axis within the extent of the curve $\Gamma$, the minimal and maximal values of $z$, denoted $z_{\min}(x)$ and $z_{\max}(x)$, were identified.

At each such position $x$, we defined a planar elliptical cross-section in the $yz$-plane, centred at
\begin{equation}
    z_m(x) = \tfrac{1}{2}\left(z_{\min}(x) + z_{\max}(x)\right).
\end{equation}
The semi-major axis of the ellipse was set to
\begin{equation}
    a(x) = \tfrac{1}{2}\left(z_{\max}(x) - z_{\min}(x)\right),
\end{equation}
and the semi-minor axis (along the $y$-direction) was defined as
\begin{equation}
    b(x) = \varepsilon a(x),
\end{equation}
with $\varepsilon = 0.6$, based on experimental observations, i.e. $\varepsilon \approx W/S$. A lattice point $(x, y, z)$, representing the centre of a sphere, was included in the body if it satisfied the condition
\begin{equation}
\left(\frac{y}{b(x)}\right)^2 + \left(\frac{z - z_m(x)}{a(x)}\right)^2 \leq 1,
\qquad \text{for } x \in \Gamma.
\end{equation}
This condition yields a shape with a surface given by the orange mesh (see figure~\ref{fig:surf_mesh}).

To introduce asymmetries characteristic of the studied geometry, two additional filters were applied. First, all lattice points lying within either of two cylindrical holes with base in the $xz$-plane were excluded. These holes had radius $r = 0.4\,s$ and were centered at $(x_c, z_c) = (\pm 1.823\,s,\ -0.548\,s)$. Figure~\ref{fig:surf_mesh} shows the surface of the holes, represented by the red mesh. Second, for each accepted cross-section, the admissible range of $y$ was further restricted by the inequality
\begin{equation}
y_{-}(x, z) \leq y \leq y_{+}(x, z), \qquad \text{where} \qquad
y_{\mp}(x, z) = \pm \tfrac{7}{20} \varepsilon z \mp \tfrac{19}{40} b(x).
\end{equation}
This condition corresponds to those spheres whose centres lie between the two green planes shown in figure~\ref{fig:surf_mesh}.

The resulting bead-based model of the crescent moon, illustrated in figure~\ref{shapes_beads_moon}, consists of 5513 spheres and corresponds to a scale factor of $s = 6.2$.

\bibliographystyle{unsrt}

\bibliography{references.bib}

\end{document}